\documentclass[%
preprint,
superscriptaddress,
 amsmath,amssymb,
 aps,
prb,
floatfix,
]{revtex4-2}

\usepackage{graphicx}
\usepackage{dcolumn}
\usepackage{bm}
\usepackage[colorlinks=true]{hyperref}
\usepackage[compat=1.1.0]{tikz-feynhand}
\usepackage[normalem]{ulem}

\begin{document}

\preprint{APS/123-QED}

\title{
    Effects of the pseudogap and the Fermi surface on the rapid
    Hall-coefficient changes in cuprates
}

\author{Yingze Su}%
\thanks{These authors contributed equally to this work.}
\affiliation{School of Physics, Peking University, Beijing 100871, China}

\author{Hui Li}
\thanks{These authors contributed equally to this work.}
\affiliation{School of Physics, Zhejiang University, Hangzhou 310058, China}

\author{Huaqing Huang}%
\email{huaqing.huang@pku.edu.cn}
\affiliation{School of Physics, Peking University, Beijing 100871, China}

\author{Dingping Li}%
\email{lidp@pku.edu.cn}
\affiliation{School of Physics, Peking University, Beijing 100871, China}




\date{\today}

\begin{abstract}
    High-$T_c$ cuprates are characterized by strong spin fluctuations,
    which give rise to antiferromagnetic and pseudogap phases
    and may be key to the high superconducting critical temperatures observed in these materials.
    Experimental studies have revealed significant changes in the Hall coefficient $R_H$ across these phases,
    a phenomenon closely related to both spin fluctuations and changes in the Fermi surface morphology.
    Utilizing the perturbation correction to Gaussian approximation (PCGA),
    the two-dimensional (2D) square-lattice single-band Hubbard model is investigated,
    resulting in the calculation of the self-energy with a finite imaginary part due to scattering.
    The density dependence of the Hall number, $n_H=1/(qR_H)$, is also calculated.
    For small hole (or electron) doping $p$ (or $x$),
    our numerical results demonstrate that $n_H$ transitions from $p$ to $1+p$ for hole-doped systems,
    and from $-x$ to $1-x$ for electron-doped systems,
    both in agreement with experimental findings.
    Furthermore, we discuss the correlation between phase boundaries and the observed peculiar changes in the Hall number.
\end{abstract}

\maketitle
\tableofcontents

\section{Introduction}
The pseudogap of cuprates is one of the most intensely debated phenomena in the studies of high-temperature superconductors \cite{Vedeneev_2021}.
Some researchers consider it a distinct phase of matter \cite{varma_1997,kivelson_1998},
while others view it as a precursor of an ordered phase \cite{Sedrakyan_2010, Wu_2019}.
The complexity of the pseudogap arises from its connection to various orders,
such as charge/spin/pair density waves \cite{Hucker_2014, Wang_2015},
stripe order \cite{comin_2015}, (short-range) antiferromagnetism \cite{Baledent_2011},
electronic nematicity \cite{Cyr_2015}, etc.
These competing orders may provide crucial insights into the emergence of high-temperature superconductivity.

In the transition from the ``normal" phase (including strange metal) to the pseudogap region, significant changes in Hall number have been observed experimentally \cite{badoux_change_2016, greene_strange_2020}. In the electron-doped region, as doping $x$ decreases, the Hall number initially drops from $1-x$ to a significantly negative value, before rising back to $-x$. This anomalous phenomenon is believed to be related to Fermi surface reconstruction \cite{greene_strange_2020}.
In the hole-doped region, the dependence of the Hall number on the doping $p$ shows a drop from $1+p$ to $p$.
Utilizing the Yang-Rice-Zhang ansatz with pseudogap, Storey pointed out that this drop is associated with $\mathrm{N\acute{e}el}$ antiferromagnetism \cite{storey_hall_2016}.
Further investigations explained the emergence of short-range order
by the Hubbard model's spiral phase.
However, it still requires numerous adjustable parameters, one of which is the scattering rate \cite{Eberlein_2016, mitscherling_longitudinal_2018}.

In this study, we employ the single-band Hubbard model to calculate the Hall coefficient across a wide range of densities.
Firstly, we attribute the dominant contribution to scattering rates to Coulomb repulsive interactions,
and calculate finite scattering rates using two-loop perturbation correction.
Subsequently, response theory is employed to obtain the conductivity and Hall number \cite{Voruganti_conductivity_1992}.
Moreover, for both hole-doped and electron-doped scenarios,
this unified description achieves behavior consistent with experimental observations under moderate doping levels.
Notably, short-range correlations and the Fermi surface both play important roles
in the abrupt changes observed in the Hall number.

This paper is structured as follows. In Sec.~\ref{chaps:formalism},
we introduce the antiferromagnetic phase and its perturbation correction for the Hubbard model. Subsequently, we provide an overview of the Fermi surface evolution with respect to density, and outline the calculation method for the Hall number. In Sec.~\ref{chaps:results}, we analyze the phase regions based on the Fermi surface, investigate the variation of the scattering rate with momentum points, and explore the changes in the Hall number across a broad range of electron filling.
Finally, Sec.~\ref{chaps:conclusion} concludes our study.

\section{Formalism\label{chaps:formalism}}
\subsection{Model and methodology}

We start with a single-band Hubbard model on the 2D square lattice.
The Hamiltonian is
\begin{equation}
    \hat H=-\sum_{\langle i,j\rangle}\sum_{\sigma}\left(
    t_{ij}\hat c^\dagger_{i\sigma}\hat c_{j\sigma}+h.c.
    \right)+U\sum_i\hat n_{i\uparrow}\hat n_{i\downarrow},
    \label{eq:hal0}
\end{equation}
where $t_{ij}$ denotes the hopping amplitude between lattice site $i$ and $j$ and $U$ represents the strength of the on-site Coulomb interaction. $\hat c^\dagger,\hat c$ are electron creation and annihilation operators, respectively. The index $\sigma=\uparrow,\downarrow$ describes the spin orientation. The hopping amplitudes have been investigated
theoretically using density-functional-theory(DFT)
calculations \cite{andersen_lda_nodate,pavarini_band-structure_2001, Imada_ab_initio_2022}, and experimentally by angle-resolved photoemission spectroscopy (ARPES) \cite{Shen_electronic_1995, Borisenko_joys_2000}.
Nearest, second nearest, and third nearest neighbor hopping $t,t',t''$ are usually taken into account and the former is adopted as the unit of energy. For $\mathrm{La}_{2-x}\mathrm{Sr}_{x}\mathrm{CuO}_{4}$ ($\mathrm{LSCO}$), $t'/t\sim-0.1$ while for $\mathrm{YBa}_2\mathrm{Cu}_3\mathrm{O}_y$ ($\mathrm{YBCO}$) and $\mathrm{Bi_2Sr_2CaCu_2O_{8+\delta}}$, $t'/t\sim-0.3$.
In our calculations, we take $t'=-0.25t, t''=0.10t$ and a pretty strong $U=6t$, near the typical values presented in literature \cite{mitscherling_longitudinal_2018, Huang_strange_2019}.
Taking the lattice constant $a$ as the unit of length, and the energy dispersion without $U$ (i.e., $U=0$) is
\begin{equation}
    \begin{aligned}
        \epsilon(\vec k)
        = & -2t\left(\cos k_x+\cos k_y\right)
        +4|t'|\cos k_x\cos k_y                     \\
          & -2t''\left(\cos 2k_x+\cos 2k_y\right).
        \label{eq:dispersion}
    \end{aligned}
\end{equation}

\begin{figure}
    \includegraphics[width=\linewidth]{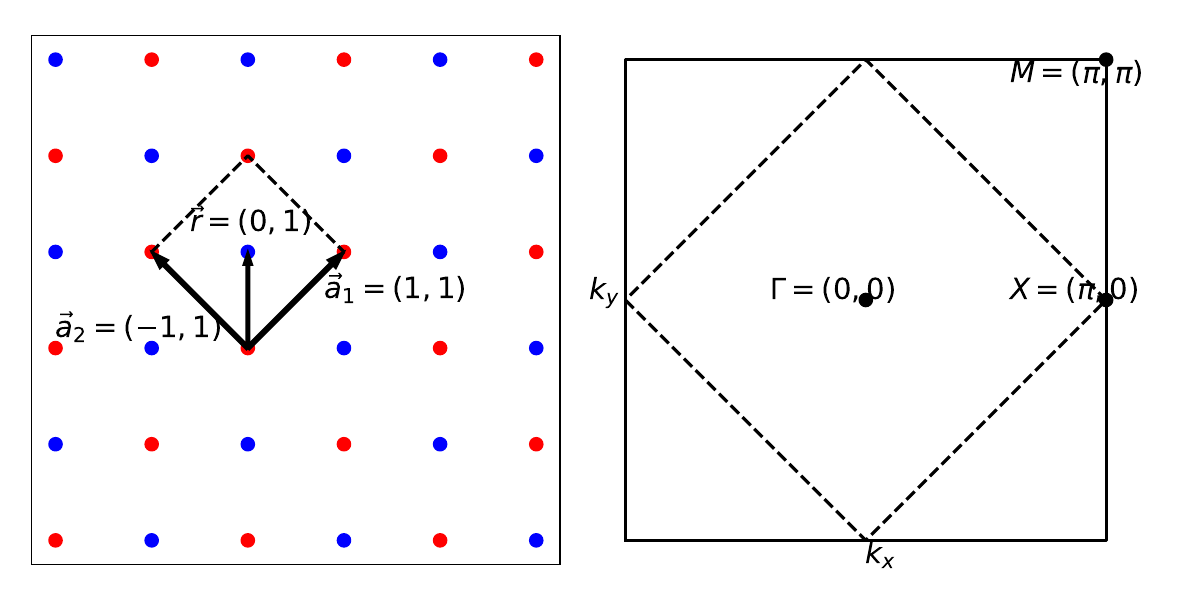}
    \caption{\label{fig:1}
        The supercell and corresponding 1st Brillouin Zone(BZ).
        Left: adjacent sites are divided into $A$ parts(red) and $B$ parts(blue).
        The lattice vectors are $\vec a_1=(1,1),\vec a_2=(-1,1)$.
        Right: the area of Folded BZ(dashed line) is half of that of the original BZ(solid line).
        The reciprocal lattice vectors are $\vec k_1=(\pi,\pi)$ and $\vec k_2=(-\pi,\pi)$}
\end{figure}

The Hartree-Fock approximation reveals that the Hubbard model exhibits a rich phase diagram \cite{Igoshev_Incommensurate_2010, Laughlin_Hartree-Fock_2014, Scholle_comprehensive_2023},
considering both magnetic and charge fluctuations.
Given the prominence of antiferromagnetic (AF) fluctuations near half-filling,
one may expect the magnetic phase transition will capture its key features.
For simplicity, we employ a paramagnetic(PM)-AF transition.
Both phases can be described in a unified form \cite{Kao_Unified_2023}.
When adjacent sites are considered separately,
the system exhibits a $\sqrt{2}\times\sqrt{2}$ super-lattice with lattice vectors $\vec a_1=(1,1)$, and $\vec a_2=(-1,1)$.
Within each super-cell, there exist two sites, denoted as $\vec r^A=(0,0)$ and $\vec r^B=(0, 1)$.
The first Brillouin zone is folded accordingly as shown in Figure~\ref{fig:1}.
The kinetic term of the Hamiltonian written in basis $\{\psi^A_\uparrow, \psi^B_\uparrow, \psi^A_\downarrow, \psi^B_\downarrow\}$ is block diagonal.
\begin{equation}
    \hat H_{0,\uparrow}(\vec k)=\begin{bmatrix}
        \epsilon_1(\vec k)                              & \epsilon_2(\vec k)e^{+\mathrm i\varphi(\vec k)} \\
        \epsilon_2(\vec k)e^{-\mathrm i\varphi(\vec k)} & \epsilon_1(\vec k)
    \end{bmatrix},
\end{equation}
\begin{equation}
    \begin{gathered}
        \epsilon_1(\vec k)=4|t'|\cos k_x\cos k_y
        -2t''\left(\cos 2k_x+\cos 2k_y\right),\\
        \epsilon_2(\vec k)
        =-2t\left(\cos k_x+\cos k_y\right),\\
        \varphi(\vec k)=\vec k\cdot\vec r^B=k_y.
    \end{gathered}
    \label{eq:dispersion AB}
\end{equation}
The Hartree-Fock Green's functions $g$ can be expressed in terms of the occupation numbers $n^A_\uparrow,n^A_\downarrow, n^B_\uparrow,n^B_\downarrow$ from Dyson-Schwinger equation,
\begin{equation}
    g^{-1}_\upuparrows(k)
    =\begin{bmatrix}
        \mathrm i\omega_n+\mu-Un^A_\downarrow
         & 0 \\ 0 &
        \mathrm i\omega_n+\mu-Un^B_\downarrow
    \end{bmatrix} - H_{0,\uparrow}(\vec k).
    \label{eq:g hf}
\end{equation}
For the AF phase, they are related to each other as $n^A_\uparrow=n^B_\downarrow, n^A_\downarrow=n^B_\uparrow$;
for the PM phase, all occupation numbers are equal.
Self-consistent equations are derived from the Matsubara sum $\frac{1}{\beta N}\sum_k g^{AA}_{\upuparrows}(k)=n^A_\uparrow$,
where $k=(\mathrm i\omega_n,\vec k)$,
and $\beta$ and $N$ denote the inverse of the temperature
and the number of sites, respectively.

Scattering due to interactions plays a pivotal role in the transport properties of cuprates,
leading us to compute the self-energy with a finite imaginary part at the two-loop level.
Making use of standard perturbation theory,
we obtain the perturbation correction to Gaussian approximation (PCGA)
\cite{Kao_Unified_2023}.
The leading term of the
self-energy $\Sigma_\upuparrows(k)$ turns out to be
\begin{equation}
    \Sigma^{ab}_\upuparrows(k)
    =-\frac{U^2}{\beta^2N^2}\sum_{q_1,q_2}
    g^{ba}_\downdownarrows(q_1+q_2-k)
    g^{ab}_\upuparrows(q_1)
    g^{ab}_\downdownarrows(q_2),
    \label{eq:self-energy}
\end{equation}
where $a,b\in\{A,B\}$. We would see that the real-frequency self-energy $\Sigma_\upuparrows(\omega,\vec k)$ does have a spatially varying imaginary part.
In Appendix \ref{ap:next_order_self_energy},
we will provide a brief derivation of the lowest non-trivial order self-energy
and estimate the magnitude of higher-order corrections.
Furthermore, the Green's function $G$ under PCGA is
\begin{equation}
    G_{\upuparrows}(k)=g_{\upuparrows}(k)
    +g_{\upuparrows}(k)\Sigma_{\upuparrows}(k)G_{\upuparrows}(k).
    \label{eq:dyson pcga}
\end{equation}

\subsection{Fermi surface\label{sec:phase_by_fs}}
\begin{figure}
    \includegraphics[width=.7\linewidth]{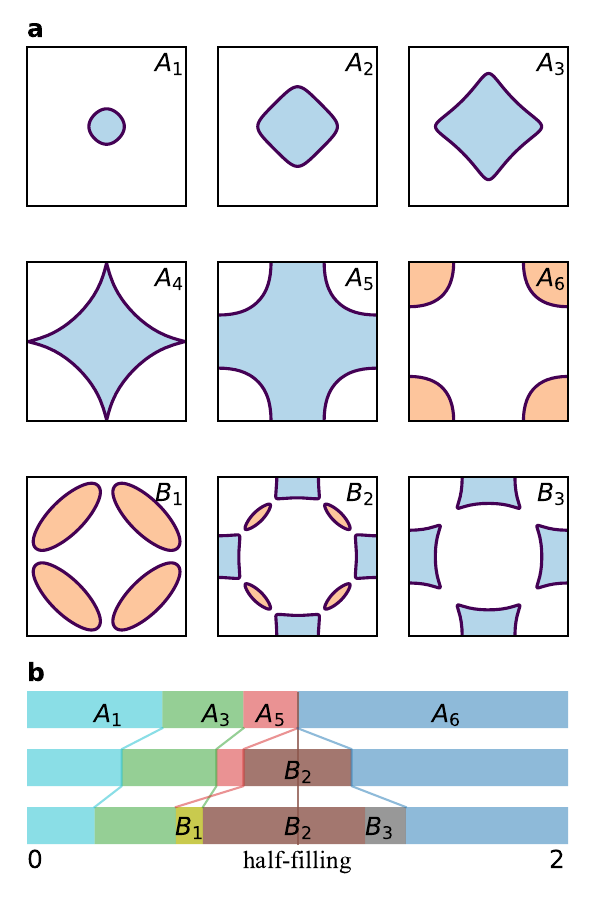}
    \caption{\label{fig:2}(a) A sketch for the FS under Hartree-Fock Approximation,
        with $(0,0)$ at the center as in Figure~\ref{fig:1}.
        The blue shaded area corresponds to electron pockets
        while the orange shaded area corresponds to hole pockets.
        $A_1\sim A_6$: PM phase, ranging
        from an extremely low density to a pretty high density.
        Around $A_2$, the FS transitions from convex to concave, and
        around $A_4$, the FS opens up.
        $B_1\sim B_3$: AF phase with finite magnetic moments
        and the electron density increases gradually.
        (b) This panel shows the phase boundary versus density at different temperatures,
        which decrease from top to bottom.
        The appearance of $B_1, B_3$ depends on the parameters we choose.
    }
\end{figure}

Various methods exist for studying phase transitions in electronic systems, one of which involves detecting the evolution of the Fermi surface (FS)
\cite{storey_hall_2016,
    armitage_angle-resolved_2003,matsui_evolution_2007,louis_remarkable_2019}.
Figure~\ref{fig:2}a shows the evolution of the FS as a function of density, which is determined using the Hartree-Fock approximation.

For the PM phase, in the dilute limit, the FS is nearly a circle, which is convex everywhere. As the electron density grows, the FS around nodal point $(q,q)$ gradually becomes concave
($A_1$ to $A_3$ in Fig.~\ref{fig:2}a).
The point that FS transitions from convex to concave is regarded as Fermi liquid starting to break down \cite{FS_Gindikin_2024}.
When anti-nodal point goes to $(\pi,0)$,
the topology of FS changes ($A_4$ in Fig.~\ref{fig:2}a).

The AF phase is more complicated
due to the presence of
finite magnetic moments $m$.
Near half-filling, electron pockets and hole pockets coexist,
as depicted in $B_2$ of Fig.~\ref{fig:2}a.
As hole/electron doping increases, the electron/hole pockets
shrink and may even vanish under certain parameter conditions.
These topological changes also indicate phase boundaries.
The boundary terminates at high temperature which
precludes the AF phase at the corresponding doping levels
(as seen between mid- and lower- temperature in Fig.~\ref{fig:2}b).

\subsection{Current and response functions\label{chaps:response}}
To calculate the Hall number, we need to evaluate longitudinal and Hall conductivity respectively. For mean-field theory with magnetic order and uniform scattering rate, there are some systematic researches \cite{Eberlein_2016, mitscherling_longitudinal_2018}. We would derive the formula in a similar way without a uniform scattering rate assumption.
Under electromagnetic field, we need to apply Peierls substitution
\cite{peierls_zur_1933, wannier_dynamics_1962, Vucice_electrical_2021} to the Hamiltonian Equation~(\ref{eq:hal0})
\begin{equation}
    t_{ij}[\vec A]=t_{ij}\exp\left(\mathrm i\left(
        \vec A_i+\vec A_j\right)\cdot(\vec r_i-\vec r_j)/2\right),
\end{equation}
where $\vec A$ is the vector potential of the electromagnetic field. The current operator $\hat j^\alpha$ and corresponding bare vertex $\gamma^\alpha$ satisfy
\begin{equation}
    \hat j^\alpha(\vec r)=\left.
    \frac{\delta\hat H[\vec A]}{\delta A_\alpha(\vec r)}
    \right|_{\vec A\equiv 0}
    =\sum_{\vec r_1,\vec r_2}\sum_{\sigma}
    c^\dagger_\sigma(\vec r_1)\gamma^\alpha(\vec r_1,\vec r_2;\vec r)
    c_\sigma(\vec r_2).
    \label{eq:vertex1}
\end{equation}
To connect (Hall) conductivity and current-correlation functions, we need to extend linear response theory up to the second order. The coefficients $\Pi$s can be expressed by either correlation functions or conductivity, serving as a bridge between them.
\begin{equation}
    \begin{aligned}
        \langle j^\alpha(\tau,\vec r)\rangle
        = & \int_0^\beta\mathrm d\tau'\ \Pi^{ab}(\tau,\vec r;\tau')A^E_b(\tau') \\
        + & \int_0^\beta\mathrm d\tau'\sum_{\vec r'}
        \Pi^{abc}(\tau,\vec r';\tau',\vec r')A^E_b(\tau')A^B_c(\vec r')         \\
        + & \text{higher order response},
    \end{aligned}
    \label{eq:current bridge}
\end{equation}
where $a,b,c\in\{x,y,z\}$ denotes spatial components, $\tau$ is the imaginary time and $\vec A(\tau,\vec r)=\vec A^E(\tau)+\vec A^B(\vec r)$.
Suppose $\vec E(\tau)=E(\tau)\hat e_x,\vec B(\vec r)=B(\vec r)\hat e_z$,
we could select appropriate gauge to make $\vec A^B$ in $y$-direction.
\begin{equation}
    \left\{\begin{aligned}
        \langle j^x(\tau,\vec r)\rangle
        =        & \int_0^\beta\mathrm d\tau\ \Pi^{xx}(\tau,\vec r;\tau')A^E(\tau')                                   \\
        \langle j^y(\tau,\vec r)\rangle
        =        & \int_0^\beta\mathrm d\tau\sum_{\vec r'} \Pi^{yxy}(\tau,\vec r;\tau',\vec r')A^E(\tau')A^B(\vec r') \\
        E(\tau)= & -\partial_\tau A^E(\tau),\quad
        B(\vec r)=\partial_x A^B(\vec r)
    \end{aligned}\right.
    \label{eq:j pi}
\end{equation}

In the language of path-integral, the expectation value of current can be expressed by action including electromagnetic field $S[\psi^*,\psi;\vec A]$
\begin{equation}
    \langle j^\alpha(\tau,\vec r)\rangle
    =-\frac{\delta}{\delta A_\alpha(\tau,\vec r)}
    \ln\int D[\psi^*,\psi]\ e^{-S[\psi^*,\psi;\vec A]}.
\end{equation}
Furthermore, take the derivative of $\vec A$ by Eq.~(\ref{eq:current bridge}).
It will be found in Sec.~\ref{chaps:number} that conductivities are only related to derivative of $\Pi^{xx}$ and $\Pi^{yxy}$ as Eq.~(\ref{eq:Kubo}), any terms with $\delta(\tau,\tau')$ or $\delta(r_x,r'_x)$ could be dropped.
\begin{equation}
    \begin{aligned}
        \Pi^{xx}(\tau,\vec r;\tau')
        =                & \sum_{\vec r'}\left\langle j^x(\tau,\vec r)j^x(\tau',\vec r')\right\rangle_c
        +\left\langle\frac{\delta j^x(\tau,\vec r)}{\delta A^E(\tau')}\right\rangle
        \to\sum_{\vec r'}\left\langle j^x(\tau,\vec r)j^x(\tau',\vec r')\right\rangle_c,                \\
        \Pi^{yxy}(\tau,\vec r;\tau',\vec r')
        =                & \int\mathrm d\tau''\sum_{\vec r''}\ \left\langle
        j^y(\tau,\vec r)j^x(\tau',\vec r'')j^y(\tau'',\vec r')
        \right\rangle_c
        +\int\mathrm d\tau''\ \left\langle
        \frac{\delta j^y(\tau,\vec r)}{\delta A^E(\tau')}
        j^y(\tau'',\vec r')
        \right\rangle_c                                                                                 \\
        +\sum_{\vec r'}  & \left\langle
        \frac{\delta j^y(\tau,\vec r)}{\delta A^B(\vec r')}
        j^x(\tau',\vec r'')
        \right\rangle_c
        +\sum_{\vec r''}\ \left\langle
        j^y(\tau,\vec r)\frac{\delta j^x(\tau',\vec r'')}{\delta A^B(\vec r')}
        \right\rangle_c
        +\left\langle\frac{\delta^2 j^y(\tau,\vec r)}
        {\delta A^B(\vec r')\delta A^E(\tau')}\right\rangle                                             \\
        \to\int\mathrm d & \tau''\sum_{\vec r''}\ \left\langle
        j^y(\tau,\vec r)j^x(\tau',\vec r'')j^y(\tau'',\vec r')
        \right\rangle_c
        +\sum_{\vec r''}\ \left\langle
        j^y(\tau,\vec r)\frac{\delta j^x(\tau',\vec r'')}{\delta A^B(\vec r')}
        \right\rangle_c.
    \end{aligned}
    \label{eq:pi}
\end{equation}
Here $\langle\cdot\rangle_c$ denotes connected correlation functions. 
The lowest order of $\Pi^{yxy}$ consists of $3$ Feynman diagrams as Fig.~\ref{fig:3}.
Bare vertices are derivatives of Hamiltonian like Eq.~(\ref{eq:vertex1})
\begin{equation}
    \begin{gathered}
        \left.\frac{\delta\hat H[\vec A]}{\delta A_\alpha(\vec r)}
        \right|_{\vec A\equiv 0}
        =\sum_{\vec r_1,\vec r_2}\sum_{\sigma}
        c^\dagger_\sigma(\vec r_1)\gamma^\alpha(\vec r_1,\vec r_2;\vec r)
        c_\sigma(\vec r_2),\\
        \left.\frac{\delta^2\hat H[\vec A]}{\delta A_\alpha(\vec r)\delta A_\beta(\vec r)}
        \right|_{\vec A\equiv 0}
        =\sum_{\vec r_1,\vec r_2}\sum_{\sigma}
        c^\dagger_\sigma(\vec r_1)\gamma^{\alpha\beta}(\vec r_1,\vec r_2;\vec r)
        c_\sigma(\vec r_2).
    \end{gathered}
\end{equation}

In momentum space, there are two entries in $\gamma^\alpha,\gamma^{\alpha\beta}$,
and we use underline to emphasize that sublattice label $A,B$
\begin{equation}
    \begin{aligned}
        \underline\gamma^{ab}(\vec k_1,\vec k_2)
        = &                                                
        \sum_{\vec r_1\in a,\vec r_2\in b}
        \gamma(\vec r_1,\vec r_2;\vec r)\times             \\
          & e^{-\mathrm i\vec k_1\cdot(\vec r_1-\vec r_2)}
        e^{-\mathrm i\vec k_2\cdot(\vec r_1-\vec r)},
    \end{aligned}
\end{equation}
where we omit superscripts for direction.
We find these two-entries $\underline\gamma$s could be expressed by one-entry $\underline\gamma$s,
\begin{equation}
    \underline\gamma(\vec k,\vec q)
    =\frac{1}{2}\left(\underline\gamma(\vec k)+\underline\gamma(\vec k+\vec q)\right)
    \simeq\underline{\gamma}(\vec k+\vec q/2),
\end{equation}
\begin{equation}
    \underline\gamma^x(\vec k)=\begin{bmatrix}
        \partial_{k_x}\epsilon_1(\vec k)                              &
        e^{+\mathrm i\varphi(\vec k)}\partial_{k_x}\epsilon_2(\vec k)   \\
        e^{-\mathrm i\varphi(\vec k)}\partial_{k_x}\epsilon_2(\vec k) &
        \partial_{k_x}\epsilon_1(\vec k)
    \end{bmatrix},
\end{equation}
\begin{equation}
    \underline\gamma^{xy}_B(\vec k)=\begin{bmatrix}
        \partial_{k_x}\partial_{k_y}\epsilon_1(\vec k)                              &
        e^{+\mathrm i\varphi(\vec k)}\partial_{k_x}\partial_{k_y}\epsilon_2(\vec k)   \\
        e^{-\mathrm i\varphi(\vec k)}\partial_{k_x}\partial_{k_y}\epsilon_2(\vec k) &
        \partial_{k_x}\partial_{k_y}\epsilon_1(\vec k)
    \end{bmatrix},
\end{equation}
$\epsilon$ and $\varphi$ are defined in Eq.~(\ref{eq:dispersion AB}),
and expression for $\underline\gamma^y$ is similar to $\underline\gamma^x$.
Although the symbol $\underline\gamma$ is employed to denote two distinct entities,
the potential for ambiguity is mitigated by explicitly
writing out the underlying entries.

Our supercells with $A,B$ sites break translation invariance, so $\Pi$s in momentum space should be symmetrized as
\begin{gather}
    \Pi^{xx}(\omega,\vec k)=
    \sum_{\vec r}
    e^{-\mathrm i\vec k\cdot\vec r}\Pi^{xx}(\omega,\vec r),\\
    \Pi^{yxy}(\omega,\vec k)=
    \sum_{\vec r_1,\vec r_2}
    e^{-\mathrm i\vec k\cdot(\vec r_1-\vec r_2)}
    \Pi^{yxy}(\omega;\vec r_1,\vec r_2).
    \label{eq:pi sym}
\end{gather}
Finally, Eqs.~(\ref{eq:pi},\ref{eq:pi sym}) could be expressed by Green's functions and vertex $\underline\gamma$
as Eq.~(\ref{eq:pi green}) in Appendix~\ref{ap:numerical}.

\usetikzlibrary{math}
\tikzmath{\r1=1.7; \r2=2.3;}
\begin{figure}[ht]
    \centering
    \begin{tikzpicture}
        \begin{feynhand}
            \vertex [dot](a) at (0, \r1) {};
            \vertex [dot](b) at (-0.866 * \r1, -0.5 * \r1) {};
            \vertex [dot](c) at (+0.866 * \r1, -0.5 * \r1) {};
            \propag [fermion] (a) to (b);
            \propag [fermion] (b) to (c);
            \propag [fermion] (c) to (a);
            \node at (0.4 * \r1, \r1) {$(\tau,\vec r,\gamma^y)$};
            \node at (-0.6 * \r1, -0.7 * \r1) {$(\tau'',\vec r',\gamma^y)$};
            \node at (+0.6 * \r1, -0.7 * \r1) {$(\tau',\vec r'',\gamma^x)$};

            \vertex (v1) at (0, \r2);
            \vertex (v2) at (-0.866 * \r2, -0.5 * \r2);
            \vertex (v3) at (+0.866 * \r2, -0.5 * \r2);
            \propag [sca] (v1) to (a);
            \propag [sca] (v2) to (b);
            \propag [sca] (v3) to (c);
        \end{feynhand}
    \end{tikzpicture}
    \quad
    \begin{tikzpicture}
        \begin{feynhand}
            \vertex [dot](a) at (0, \r1) {};
            \vertex [dot](b) at (-0.866 * \r1, -0.5 * \r1) {};
            \vertex [dot](c) at (+0.866 * \r1, -0.5 * \r1) {};
            \propag [fermion] (a) to (c);
            \propag [fermion] (c) to (b);
            \propag [fermion] (b) to (a);
            \node at (0.4 * \r1, \r1) {$(\tau,\vec r,\gamma^y)$};
            \node at (-0.6 * \r1, -0.7 * \r1) {$(\tau'',\vec r',\gamma^y)$};
            \node at (+0.6 * \r1, -0.7 * \r1) {$(\tau',\vec r'',\gamma^x)$};

            \vertex (v1) at (0, \r2);
            \vertex (v2) at (-0.866 * \r2, -0.5 * \r2);
            \vertex (v3) at (+0.866 * \r2, -0.5 * \r2);
            \propag [sca] (v1) to (a);
            \propag [sca] (v2) to (b);
            \propag [sca] (v3) to (c);
        \end{feynhand}
    \end{tikzpicture}
    \quad
    \begin{tikzpicture}
        \begin{feynhand}
            \vertex [dot](a) at (-\r1, 0) {};
            \vertex [crossdot](b) at (+\r1, 0) {};
            \propag[fermion] (a) to[out=60, in=120] (b);
            \propag[fermion] (b) to[out=-120, in=-60] (a);
            \node at (-1.3 * \r1, -0.2 * \r1) {$(\tau,\vec r,\gamma^y)$};
            \node at (+1.3 * \r1, -0.2 * \r1) {$(\tau',\vec r',\gamma^{xy}_B)$};

            \vertex (v1) at (-\r2, 0);
            \vertex (v2) at (+\r2, \r1 - \r2);
            \vertex (v3) at (+\r2, \r2 - \r1);
            \propag [sca] (v1) to (a);
            \propag [sca] (b) to (v2);
            \propag [sca] (b) to (v3);
        \end{feynhand}
    \end{tikzpicture}
    \caption{\label{fig:3}Leading Feynman diagrams for $\Pi^{yxy}(\tau,\vec r;\tau',\vec r')$.
        At each vertex, there should be an imaginary time $\tau$,
        a spatial coordinate $\vec r$ and a bare vertex $\gamma$.
        The internal indices $\tau'',\vec r''$ should be integrated out.
    }
\end{figure}

\subsection{Hall number\label{chaps:number}}
The longitudinal and Hall conductivity $\sigma,\sigma_\mathrm{Hall}$
are described by the current response to homogeneous static electric and magnetic fields,
\begin{equation}
    j^x=\sigma E,\quad
    j^y=\sigma_\mathrm{Hall}BE.
\end{equation}
And we get frequency-dependent conductivity
expressed by $\Pi$s in Eq.~(\ref{eq:j pi}) and Eq.~(\ref{eq:pi sym})
as proved in \cite{Voruganti_conductivity_1992}
\begin{equation}
    \begin{gathered}
        \sigma(\omega)=\frac{1}{\mathrm i\omega}\left(
        \Pi^{xx}(\omega,\vec k=\vec 0)-(\omega\to0)
        \right),\\
        \sigma_\mathrm{Hall}(\omega)
        =\frac{1}{\omega}\frac{\partial}{\partial k_x}
        \left(\Pi^{yxy}(\omega,\vec k=\vec 0)-(\omega\to 0)\right).
    \end{gathered}
    \label{eq:Kubo}
\end{equation}
By definition, the static limit is defined as $\sigma=\sigma(\omega=0), \sigma_\mathrm{Hall}=\sigma_\mathrm{Hall}(\omega=0)$.

Once $\Pi(\tau,\tau')$ is obtained, we still need some analytical continuation techniques, or proxies instead. For longitudinal conductivity, there are two well-known proxies $\sigma_1=\beta^2\Lambda(\beta/2)/\pi$, $\sigma_2=(2\pi\Lambda(\beta/2))^2/\Lambda''(\beta/2)$, where $\Lambda(\tau)=\Pi^{xx}(\tau,\vec 0)$ \cite{Samuel_Superconductivity_2017,Huang_strange_2019}.
There are also similar proxies for the Hall conductivity and the Hall number in previous researches \cite{Huang_Numerical_2021}. However, many of them are not unbiased with finite scattering rates, and only work at low temperatures. We want to establish/choose unbiased proxies which are consistent with the Drude theory.

The equation of motion for a semi-classical particle in an electromagnetic field can be expressed as
\begin{equation}
    \frac{\mathrm d\vec p}{\mathrm dt}
    =q(\vec E+\vec v\times\vec B)
    -2\eta m\vec v,
\end{equation}
where $\vec p=m\vec v$, and $\eta>0$ is the relaxing rate.
Our proxies should not depend on $\eta$.
Time-dependent electric field $\vec E=Ee^{-\mathrm i\omega t}\hat e_x$
and static magnetic field $B=B\hat e_z$ are applied to the system.
The conductivity is defined as the response function in the weak-field limit
\begin{equation}
    nqv_x=\sigma(\omega)Ee^{-\mathrm i\omega t},\quad
    nqv_y=\sigma_\mathrm{Hall}(\omega)BEe^{-\mathrm i\omega t},
\end{equation}
where $n$ denotes the density of the particle. By solving these equations, we obtain the conductivity
\begin{equation}
    \sigma(\omega)=\frac{1}{2\eta-\mathrm i\omega}\frac{nq^2}{m},\quad
    \sigma_\mathrm{Hall}(\omega)=\frac{1}{(2\eta-\mathrm i\omega)^2}\frac{nq^3}{m^2}.
    \label{eq:sigma with eta}
\end{equation}
Therefore, the dependence of correlation functions $\Pi$ on $\mathrm i\omega_n$ is derived from Eq.~(\ref{eq:Kubo})
\begin{equation}
    \begin{gathered}
        \Pi^{xx}(\mathrm i\omega_n,\vec k=\vec 0)
        =-\frac{\omega_n}{2\eta+\omega_n}2\eta\sigma
        +\mathrm{const},\\
        \frac{\partial}{\partial k_x}
        \Pi^{yxy}(\mathrm i\omega_n,\vec k=\vec 0)
        =\frac{\mathrm i\omega_n}{(2\eta+\omega_n)^2}
        (2\eta)^2\sigma_\mathrm{Hall}
        +\mathrm{const}.
    \end{gathered}
    \label{eq:pis omega}
\end{equation}
The Hall coefficient $R_H=\sigma_\mathrm{Hall}/\sigma^2$ has an unbiased estimation by $\Pi$s on non-zero Matsubara frequencies $\mathrm i\omega_n$, and it is indeed one of the M-type in \cite{Huang_Numerical_2021}.
\begin{equation}
    R_H(\mathrm i\omega_n)
    =\left.\frac{\mathrm i\omega_n}{[\Pi^{xx}(\mathrm i\omega_n,\vec k)]^2}
    \frac{\partial\Pi^{yxy}(\mathrm i\omega_n,\vec k)}{\partial k_x}\right|_{\vec k=0}.
    \label{eq:Rh}
\end{equation}
Other candidate proxies are presented in Appendix \ref{ap:proxies}.
In the Sec.~\ref{sec:hall},
we opt to present the variations in the Hall number $n_H=1/(qR_H)$
(here electron carries charge $q=1$)
rather than directly exhibiting $R_H$.
To maintain the coherence and readability of the main text,
we have compiled the intricate details of the numerical calculations in Appendix~\ref{ap:numerical}.

In the Drude theory, the right side of Eq.~(\ref{eq:Rh}) is independent of $\mathrm i\omega_n$,
which may not be valid in general theory.
However, for low enough temperature, $R_H$ on some lowest frequencies varies slowly, allowing us to estimate the zero-frequency limit safely.

\section{Results\label{chaps:results}}
\subsection{Phase diagram}

As mentioned in Sec.~\ref{sec:phase_by_fs}, there are two typical characteristic points for the PM phase and two for the AF phase (if it exists). Phase boundary with the lowest density $n$ refers to FS changes from convex $A_1$ to concave $A_3$, i.e. there exists a momentum point $(p,p)$ so that shifted chemical potential $\tilde\mu=\mu-Un/2$ satisfies $\epsilon(p,p)=\tilde\mu$ and
$\frac{\partial^2}{\partial q^2}\epsilon(p+q, p-q)=0$. Other phase boundary $A_3$ to $A_5$ satisfies $\epsilon(0,\pi)=\tilde\mu$ 
and $B_2$ to $B_3$ satisfies $\epsilon(\pi/2,\pi/2)=\tilde\mu$. By utilizing Eq.~(\ref{eq:dispersion}), we can solve these conditions respectively to yield
\begin{gather}
    \tilde\mu_1=\frac{|t'|^2}{t''}+t(|t'|+6t'')\frac{t-
        \sqrt{t^2+64t''(|t'|+2t'')}
    }{32t''^2}, \\
    \tilde\mu_2 = -4(|t'|+t''),\\
    \tilde\mu_3=4t''.
\end{gather}
When $t>2|t'| > 4t''>0$, which is satisfied by parameters we use, the order $\tilde\mu_1<\tilde\mu_2<0<\tilde\mu_3$ holds. Furthermore, $\tilde\mu_1$ corresponds to Fermi liquid starting to break down, denoted by subscript ``$\mathrm{FL}$". $\tilde\mu_2,\tilde\mu_3$ are denoted by ``$\mathrm{topo}$"
to emphasize topological changes in FS.

\begin{figure}
    \includegraphics[width=.9\linewidth]{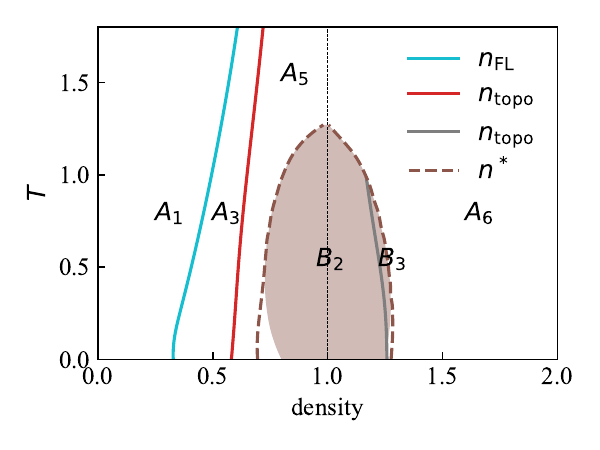}
    \caption{\label{fig:4}
        Phase diagrams for PCGA.
        When density is around or lower than $n_\mathrm{FL}$,
        the system is definitely a Fermi liquid.
        When the density is around $n_\mathrm{topo}$ and at half-filling $n=1$,
        the effective carrier type switches frombetween electrons and holes.
        In the shaded area,
        the PM phase is no longer stable,
        leading to a pseudogap.
        The FS for each region is labeled as $A_n$ and $B_n$ fromas in Fig.~\ref{fig:2}.
    }
\end{figure}

For Hartree-Fock Approximation, $\tilde\mu$ on phase boundaries does not vary with temperature,
density would vary as shown in Fig.~\ref{fig:4}.
Moreover, AF regions, which are pseudogap indeed, should be determined.
Instability $\chi_\mathrm{sp}(\omega=0,\vec k=(\pi,\pi))$ serves as a preliminary argument.
\begin{gather}
    \chi_\mathrm{sp}(\mathrm i\omega_n,\vec k)
    =\frac{\chi_0(\mathrm i\omega_n,\vec k)}
    {1-\frac{U}{2}\chi_0(\mathrm i\omega_n,\vec k)},\\
    \chi_0(\mathrm i\omega_n,\vec k)=-2
    \int\frac{\mathrm d^2\vec q}{(2\pi)^2}
    \frac{f(\epsilon_{\vec q})-f(\epsilon_{\vec k+\vec q})}
    {\mathrm i\omega_n+\epsilon_{\vec q}-\epsilon_{\vec k+\vec q}},\\
    f(\epsilon)=\frac{1}{e^{\beta(\epsilon-\mu)}+1}.
\end{gather}
At relatively low temperatures, we find that the AF phase exists in a wider region than that of the region determined by instability. It gives estimates of the boundary of
pseudogap and normal phase.

Phases are rich for low temperatures. Near half-filling, there is an AF phase region. In the electron-doped region, the disappearance of hole pockets could happen, followed by AF-PM phase transition. In the hole-doped region, AF-PM phase transition occurs earliest, after which the topological properties and concavities of FS change successively. For the remainder of this article, we focus on a low enough temperature $T=0.1$.

\subsection{\label{sec:pseudogap}Pseudogap}
\begin{figure}
    \centering
    \includegraphics[height=.7\textheight]{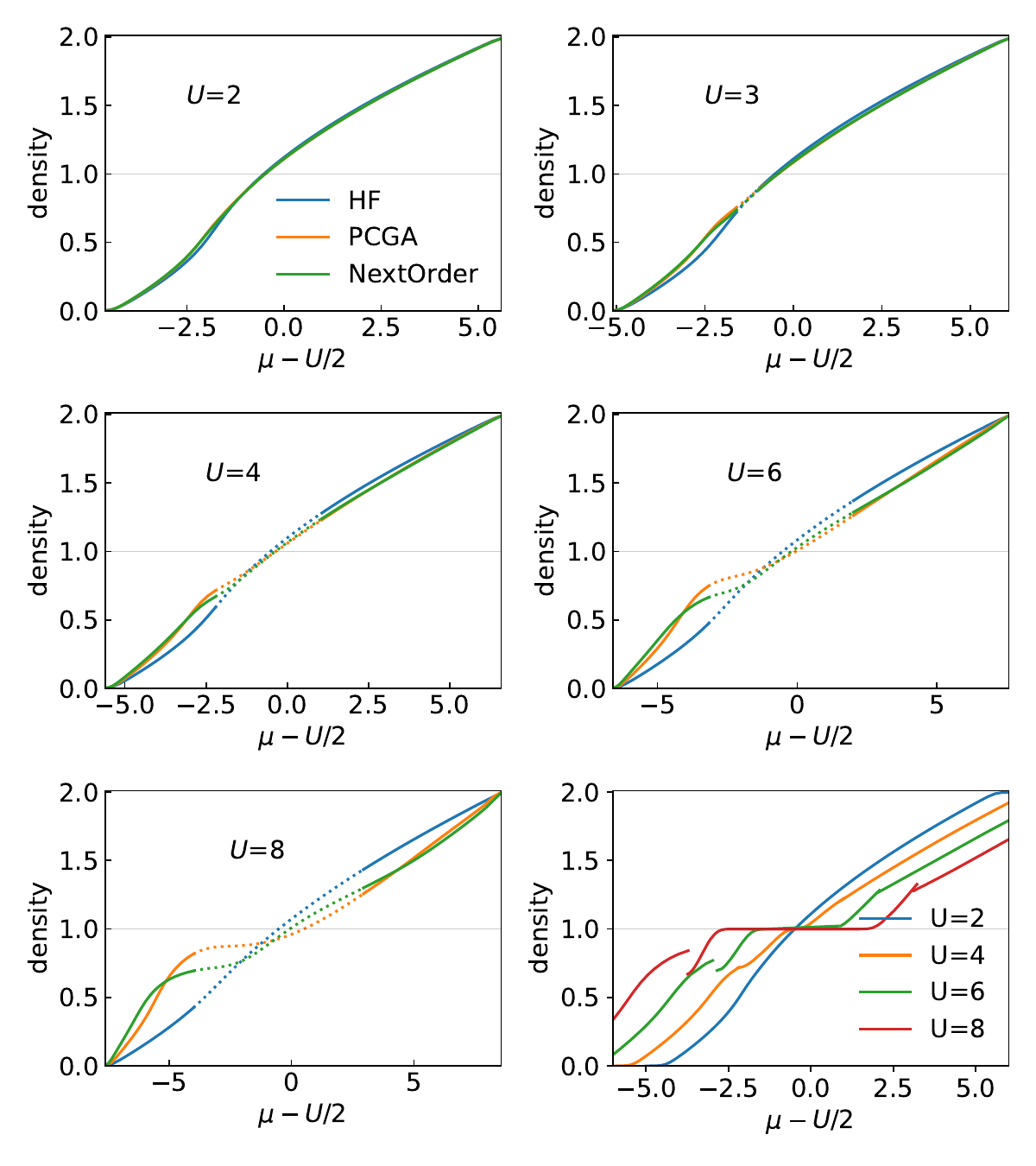}
    \caption{\label{fig:5}
        The first five figures illustrate the Hartree-Fock (HF),
        first-order correction (denoted as PCGA), and second-order correction (denoted as NextOrder) results
        for these density variations over the chemical potential $\mu$.
        The dashed lines denote the unstable PM phase.
        In contrast, the final figure delineates the existence regions of the AF phase at different $U$ values.
        Specifically, at $U=2$, the system exhibits full PM behavior across the entire region.
        As $U$ increases to 4, there is an approximate continuous transition between the PM and AF phases.
        At higher U values, such as $U=6$ and $U=8$,
        the transition between the two phases becomes markedly discontinuous.
    }
\end{figure}

It is widely acknowledged that the emergence of pseudogap is strongly contingent upon the strength of the interaction $U$.
In our study, we investigate the doping range over which pseudogaps persist at
the fixed temperature $T=0.1$ as a function of $U$
in the last panel of Fig.~\ref{fig:5}.
Notably, at low values of $U\lesssim 3$, pseudogaps are not observed to form.
As $U$ reaches intermediate strengths,
PCGA yield consistent pseudogap results in both the PM and AF phases.
However, at high values of $U$,
the AF phase exhibits a broader range of pseudogaps.
We posit that the computational reliability of the crossover region at these elevated $U$ values is questionable,
suggesting a diminished confidence in the results obtained for this regime.

In our analysis, we have also estimated the effects of the next-order perturbation corrections as depicted in Fig.~\ref{fig:5},
with further details provided in Appendix \ref{ap:next_order_self_energy}.
It is observed that when $U$ is not exceedingly large,
only the lowest non-trivial order perturbation correction,
namely PCGA employed in this study,
plays a significant role.
Upon employing $U=6$,
we note a decrease in reliability near the PM to AF phase transition region;
however, the PCGA continues to exert a dominant influence in this context.

\subsection{Momentum dependent scattering rate}

The phenomenon that Hall number changes sharply over hole doping has been studied widely \cite{klebel-knobloch_transport_2023, storey_hall_2016, mitscherling_longitudinal_2018}.
Notably, existing investigations have predominantly employed a uniform scattering rate as an adjustable parameter.
While this approximation aligns reasonably well with experimental data,
we have observed that the scattering rate exhibits significant variation across momentum values.
Specifically, the longitudinal conductivity primarily depends on momentum around the Fermi surface,
whereas the Hall conductivity is influenced by momentum below the Fermi surface.

\begin{figure}
    \centering
    \includegraphics[width=.53\linewidth]{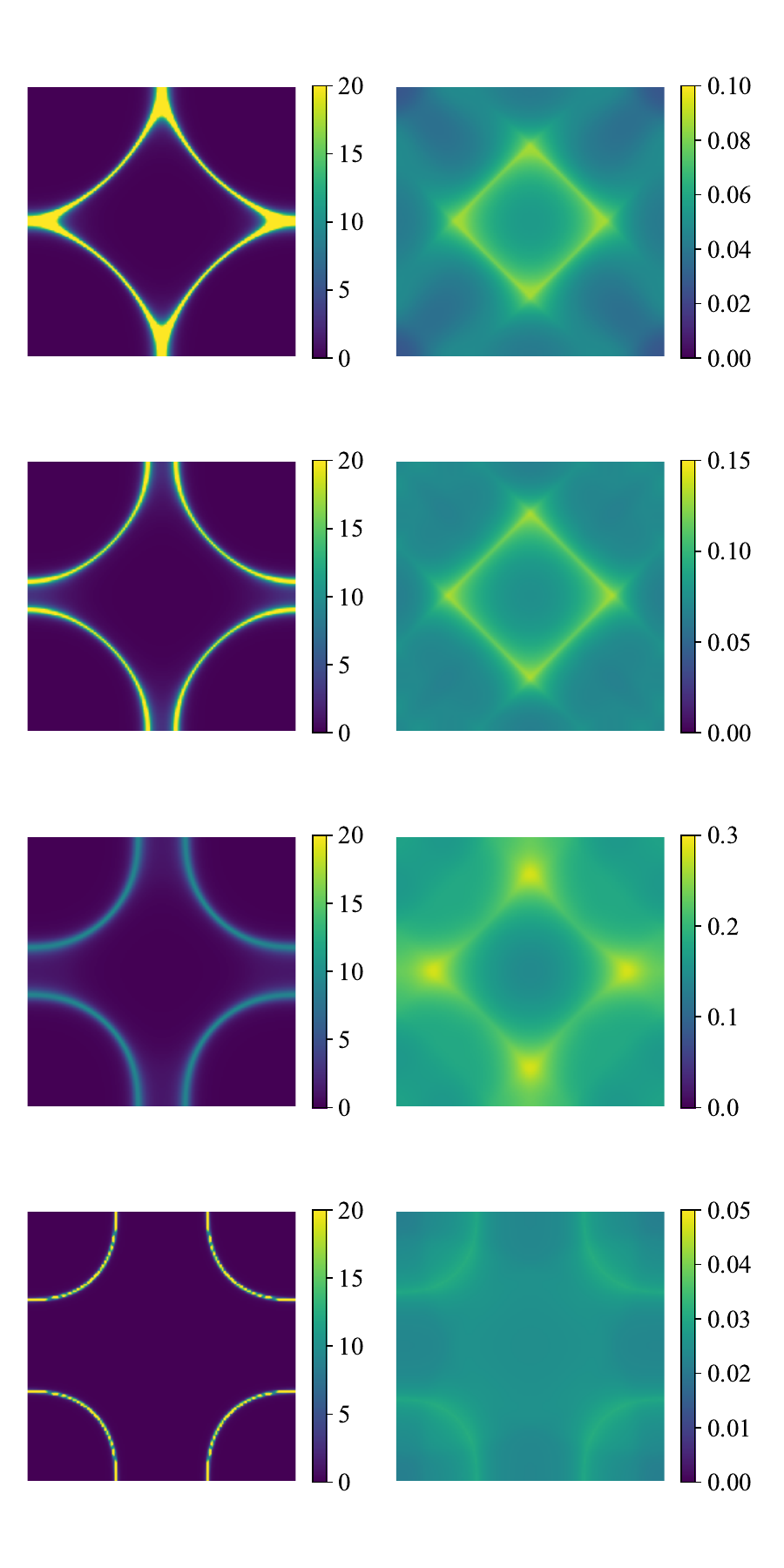}
    \caption{\label{fig:6}
        Spectral functions $A(\omega=0,\vec k)$ and the imaginary parts of self-energy $\Sigma(\omega=0,\vec k)$
        for multiple datasets.
        The spectral functions are on the left and
        the imaginary parts of the self-energy are on the right.
        From top to bottom, the densities corresponding to each dataset are approximately
        $0.56,0.67,0.77$ and $1.28$, respectively.
        The smaller the typical value of $\mathrm{Im}\Sigma$,
        the closer $A(\omega)$ approaches a Dirac-$\delta$ function near its peak,
        and consequently,
        the larger the numerical value of the peak.
        For aesthetic purposes,
        we employ the same color bar scale for $A$
        while using different scales for $\mathrm{Im}\Sigma$.
    }
\end{figure}

In Fig.~\ref{fig:6}, we present the imaginary part of the self-energy $\mathrm{Im}\Sigma$,
which represents the magnitude of scattering rate $\Gamma$,
as a function of specific points within the Brillouin zone for various doping scenarios,
including both hole-doped and electron-doped cases.
In the electron-doped region,
as depicted in the last row of Fig.~\ref{fig:6},
the shape of the spectral function and the scattering rate are broadly consistent,
indicating that the results would not significantly differ from
those obtained with a spatially uniform scattering rate.
Conversely, in the hole-doped region,
represented by the first four rows,
there is a distinct discrepancy between the peak positions of the scattering rate and the spectral function,
necessitating a more refined analysis of the spatial distribution of the scattering rate.
It is important to note that when calculating the spectral function and scattering rate from Eq.~(\ref{eq:sigam integrated out}),
we employ the identity Eq.~(\ref{eq:identity}).
We incrementally increase the size until the shapes remain roughly unchanged(largest up to $128\times128$),
although numerical precision introduces minor ripples in the details.

\subsection{\label{sec:hall}Hall number for broad-ranged density}
\begin{figure}
    \includegraphics[width=\linewidth]{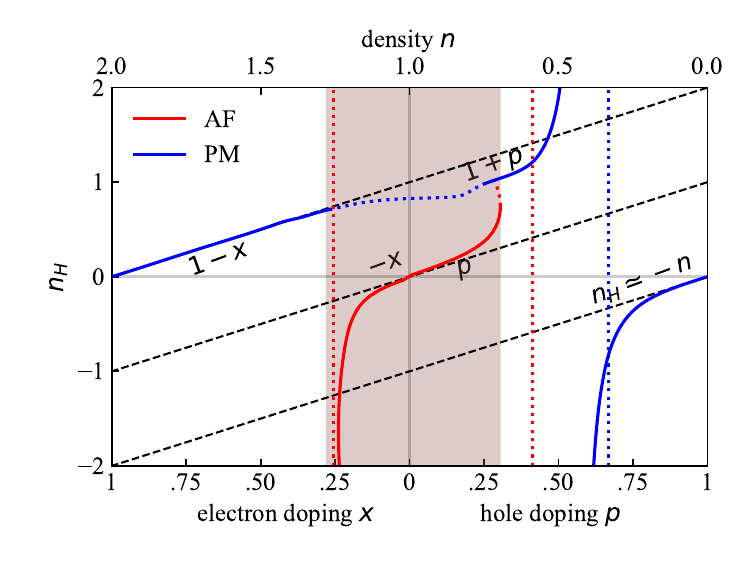}
    \caption{\label{fig:7}
        Hall number for a broad-range of density $n$.
        The hole doping concentration is denoted by $p$
        and the electron doping concentration is denoted by $x$.
        Dotted lines denote the unstable phases.
        Near the critical regions, there are some artifact phases
        due to the limitations of mean-field theory.
        Higher-order approximations will mitigate this unphysical discontinuity.
        Red Dashed lines denote changes in topological properties of FS,
        while blue dashed lines denote changes in the concavity of FS.
        The right half part shows $n_H$ transition from $p$ to $1+p$
        in the hole-doped region,
        and the left half part shows transition from $-x$ to $1-x$
        in the electron-doped region.
    }
\end{figure}

Now we consider the variation of the Hall number as a function of density.
It is pertinent to relate the Hall number, $n_H$, to the density of holes. Particularly noting that at low hole doping levels, $n_H$ remains positive and closely approaches the doping concentration $p$. Within this context, three distinct phases emerge: the paramagnetic phase, the antiferromagnetic phase,
and the enforced paramagnetic phase (with negative AF instability).

Firstly, we focus on the paramagnetic phase. In scenarios where the density is lower than $n_\mathrm{FL}$, electrons are the effective carriers, as indicated by a negative Hall number $n_H \simeq -n$. When densities increase to near $n_\mathrm{topo}$ within the hole-doped region, ``holes" become the effective carriers, with the Hall number approaching $n_H\simeq 2-n$, i.e. $1+p$ expressed by hole doping $p$. Notably, when hole doping is marginally below $p_\mathrm{topo}$, the Hall number is less than $2-n$, signifying that both electrons and holes play non-negligible roles in charge transport. Moreover, the Hall number can far exceed $2-n$ when doping is above $p_\mathrm{topo}$, consistent with experimental observations \cite{badoux_change_2016}.

The cases of hole doping and electron doping behave very differently.
In the context of hole doping, when doping is small and falls within the AF region, $n_H$ is close to the doping $p$. This is followed by a rapid but continuous ascent to $1+p$.
In the context of electron doping, $x_\mathrm{topo}$ is proximate to the boundary of PM instability. Experiments\cite{greene_strange_2020,dagan_fermi_2016}
and our results both exhibit two distinct segments: $n_H\simeq-x$  for $x<x_\mathrm{topo}$ and $n_H\simeq 1-x$ for $x>x_\mathrm{topo}$.
Around their boundary, absolute values of $n_H$ become very large.

\section{Conclusion\label{chaps:conclusion}}
We use a unified single-band Hubbard model with a non-zero next-nearest neighbor hopping amplitude $t'$ and apply the PCGA method to obtain a momentum-dependent scattering rate. We then calculate the variation of the Hall number $n_H$ across different densities. Near half-filling, a sharp transition driven by strong magnetic fluctuations is observed. In the hole-doped region, $n_H$ changes continuously from $p$ to $1+p$, while in the electron-doped region, it exhibits a discontinuous shift from $-x$ to $1-x$. Additionally, at very low electron densities $n$, there is a small interval where $n_H\simeq -n$, indicating a switch in the effective carrier type.

In addition to strong magnetic fluctuations, changes in the topology and concavity of the Fermi surface (FS) play a significant role in the unusual behavior of $n_H$. Specifically, the variations in FS topology account for the switch in the effective carrier type, reflected in the sign change of $n_H$. A convex FS typically corresponds to a Fermi liquid, which leads to certain singularities in $n_H$.

In experimental observations,
it has been noted that the pseudogap region is not only characterized by
spin fluctuations but also exhibits contributions from
the charge channel and the Cooper pair channel, as described in the literature \cite{vishik_photoemission_2018}.
Our current work, while not taking into account the influence of other order parameters,
is still capable of capturing the principal behavior of the Hall number $n_H$.
Some studies utilizing alternative methods have indicated that within the pure Hubbard model,
the pseudogap is not easily grasped in terms of contributions from other channels\cite{dong_dynamical_2019}.
Therefore, it is imperative in our subsequent work to consider electron-phonon interactions
or electron-electron nearest-neighbor interactions to better understand the contributions to the pseudogap.

We should also note that the AF phase breaks continuous symmetry, which is forbidden by the Mermin-Wagner theorem. Spontaneous symmetry breaking induces Goldstone modes, and destroys any long-range order (LRO). However, short-range order (SRO) survives and affects physical observables. By doing symmetrization in each phase-breaking direction, any LRO-related quantities vanish, and other quantities improve as SRO is taken into account 
\cite{JEVICKI_1977}.

\begin{acknowledgments}
    This work is supported by the National Natural Science Foundation of China (Grants No. 12174006 and No. 12474056 of Prof. Li’s fund) and the High-performance Computing Platform of Peking University. H.H. acknowledges the support of the National Key R\&D Program of China (No. 2021YFA1401600), and the National Natural Science Foundation of China (Grant No. 12474056). The authors are very grateful to B. Rosenstein for valuable discussions and help in numerical computations.
\end{acknowledgments}

\appendix


\section{\label{ap:proxies}
  Proxies for Hall coefficient}
There are two types of proxies, referred to as D-type and M-type, respectively.
The D-type is concerned with $\Pi$s at the imaginary time $\tau=\beta/2$,
and it is related to Matsubara frequencies by
\begin{equation}
    \Pi(\tau)=\frac{1}{\beta}\sum_{\mathrm i\omega_n}
    e^{-\mathrm i\omega_n\tau}\Pi(\mathrm i\omega_n).
\end{equation}
Due to the fluctuation-dissipation theorem, we have
\begin{equation}
    \Pi(\mathrm i\omega_n)=-\int_{-\infty}^\infty
    \frac{\mathrm d\omega}{\pi}\ \frac{
        \mathrm{Im}\Pi(\omega)
    }{\mathrm i\omega_n-\omega},
\end{equation}
Summing over $\mathrm i\omega_n$, we get
\begin{equation}
    \Pi(\tau)=\int_{-\infty}^\infty\frac{\mathrm d\omega}{\pi}
    \frac{e^{(\beta-\tau)\omega}}{e^{\beta\omega}-1}
    \mathrm{Im}\Pi(\omega),
\end{equation}
And in Eq.~(\ref{eq:Kubo}), $\mathrm{Im}\Pi(\omega)$ is replaced by the frequency-dependent conductivity $\sigma(\omega)$.
For example
\begin{equation}
    \mathrm{Im}\Pi^{xx}(\omega,\vec 0)=\omega\mathrm{Re}\sigma(\omega),
\end{equation}
\begin{equation}
    \implies\quad
    \Pi^{xx}(\tau=\beta/2,\vec 0)
    =\int_{-\infty}^\infty\frac{\mathrm d\omega}{2\pi}
    \frac{\omega}{\sinh(\beta\omega/2)}
    \mathrm{Re}\sigma(\omega).
\end{equation}
Suppose temperature $T=1/\beta$ is low enough, i.e. $T\ll\Lambda$
with $\Lambda$ representing the characteristic width of $\sigma(\omega)$,
we can integrate out
\begin{equation}
    \Pi^{xx}(\beta/2,\vec 0)
    \simeq\sigma\int_{-\infty}^\infty\frac{\mathrm d\omega}{2\pi}
    \frac{\omega}{\sinh(\beta\omega/2)}
    =\frac{\pi}{\beta^2}\sigma.
    \label{eq:pixx tau}
\end{equation}
Similarly, $\Pi^{yxy}$ satisfies
\begin{equation}
    \frac{\partial}{\partial k_x}\Pi^{yxy}(\tau,\vec 0)
    =\mathrm i\int_{-\infty}^\infty\frac{\mathrm d\omega}{\pi}
    \frac{\omega e^{(\beta-\tau)\omega}}{e^{\beta\omega}-1}
    \mathrm{Im}\sigma_\mathrm{Hall}(\omega).
\end{equation}
Since $\mathrm{Im}\sigma_\mathrm{Hall}(\omega)$ is an odd function, $\tau=\beta/2$ yields zero result,
but its derivative with respect to $\tau$ is finite,
\begin{equation}
    \frac{\partial}{\partial\tau}\frac{\partial}{\partial k_x}
    \Pi^{yxy}(\beta/2,\vec 0)
    =-\mathrm i\int_{-\infty}^\infty\frac{\mathrm d\omega}{2\pi}
    \frac{\omega^2}{\sinh(\beta\omega/2)}
    \mathrm{Im}\sigma_\mathrm{Hall}(\omega).
\end{equation}
Now we get the D-type proxy
\begin{equation}
    R^D_H=\frac{1}{\mathrm i\beta}\frac{1}{\Pi^{xx}(\beta/2,\vec 0)^2}
    \frac{\partial^2}{\partial\tau\partial k_x}
    \Pi^{yxy}(\beta/2,\vec 0).
\end{equation}
The M1-type has been derived in \ref{chaps:number}.
\begin{equation}
    R_H^{M1}=\frac{\mathrm i\omega_n}{\Pi^{xx}(\mathrm i\omega_n,\vec 0)^2}
    \frac{\partial\Pi^{yxy}(\mathrm i\omega_n,\vec 0)}{\partial k_x}.
\end{equation}
Combining Eqs.~(\ref{eq:pis omega}) and (\ref{eq:pixx tau}) ,
we find that the M2-type turns out to be
\begin{equation}
    R_H^{M2}\simeq\frac{\pi^2}{\mathrm i\omega_n\beta^4}
    \frac{\partial_{k_x}\Pi^{yxy}(\mathrm i\omega_n=2\pi\mathrm i/\beta,\vec 0)}
    {\Pi^{xx}(\tau=\beta/2,\vec 0)^2}.
\end{equation}
\section{\label{ap:ac_for_self_energy}
  Analytical continuation for the self-energy}
We start from Eq.~(\ref{eq:self-energy}), and rewritten Green's function by spectral representation
\begin{equation}
    G(\mathrm i\omega_n,\vec q)
    =\int_{-\infty}^\infty\frac{\mathrm d\omega}{2\pi}
    \frac{\mathcal A(\omega,\vec q)}{\mathrm i\omega_n-\omega}
    ,\quad
    \mathcal A(\omega,\vec q)=2\pi\delta(\omega-\xi_{\vec q}).
\end{equation}
And summing internal Matsubara frequencies out
\begin{equation}
    \Sigma(\mathrm i\omega_n,\vec k)
    =-\frac{U^2}{4N^2}\sum_{\vec q_1\vec q_2}\left[
        \frac{1}{\mathrm i\omega_n-(\xi_1+\xi_2-\xi_3)}
        \frac{\cosh\left(\beta(\xi_1+\xi_2-\xi_3)/2\right)}
        {\cosh(\beta\xi_1/2)\cosh(\beta\xi_2/2)\cosh(\beta\xi_3/2)}
        \right]
    \label{eq:sigam integrated out}
\end{equation}
Here $\xi_1=\xi_{\vec q_1},\xi_2=\xi_{\vec q_2},\xi_3=\xi_{\vec q_1+\vec q_2-\vec k}$.
Now analytical continuation is almost well-defined by substitution
\begin{equation}
    \frac{1}{\mathrm i\omega_n-\xi}\to
    \frac{1}{\omega-\xi+\mathrm i\eta}
    =\mathcal P\frac{1}{\omega-\xi}-\pi\delta(\omega-\xi)
    \label{eq:identity}
\end{equation}
For numerical simulation,
we have to take a series of small $\eta=\eta(N)$ to obtain finite imaginary part.

\section{\label{ap:numerical}
  Numerical detail}
In the main text,
specifically in Sec.~\ref{chaps:formalism},\ref{chaps:response} and \ref{chaps:number},
we present the essential formulas required for the computation of Hall numbers $n_H$.
In summary, the process can be delineated into the following steps:
\begin{enumerate}
    \item calculating the Hartree-Fock Green's function $g$ as the starting point for perturbation theory;
    \item discretizing imaginary time;
    \item computating the Green's function $G$,
    \item computating the correlation functions $\Pi$;
    \item utilizing proxies to compute the Hall number $n_H$.
\end{enumerate}
We will subsequently provide detailed numerical specifics,
along with the necessary parameters that we utilize.

\textit{Hartree-Fock Green's function $g$} is provided by Eq.~(\ref{eq:g hf}).
In fact, within the context of condition $\frac{1}{\beta N}\sum_k g^{AA}_{\upuparrows}(k)=n_\uparrow^A$,
the summation over Matsubara frequencies can be analytically given
\begin{equation}
    \begin{gathered}
        \frac{1}{N}\sum_{\vec k}\left[e^{\beta(\hat H_{0,\uparrow}(\vec k)+\Omega_{\uparrow})}+1\right]^{-1}
        =\begin{bmatrix}
            n_\uparrow^A & ??           \\
            ??           & n_\uparrow^B
        \end{bmatrix},\\
        \Omega_{\uparrow}=\begin{bmatrix}
            Un_\downarrow^B -\mu & 0                   \\
            0                    & Un_\downarrow^A-\mu
        \end{bmatrix},
    \end{gathered}
\end{equation}
where $??$ means the off-diagonal elements do not influence the solution.
We have chosen a system of size $128\times 128$ to calculate these four local densities
$n^A_\uparrow,n^A_\downarrow,n^B_\uparrow,n^B_\downarrow$,
thereby obtaining the Green's function at Matsubara frequencies.

\textit{The imaginary time discretization} is carried out
using the coherent state path integral approach as described in literature \cite{negele_quantum_2018}.
By uniformly dividing the imaginary time into $M$ segments,
the partition function for a free system is expressed as
\begin{equation}
    \begin{gathered}
        Z=\lim_{M\to\infty}\int\prod_{m=1}^M\mathrm d\bar\psi_m\mathrm d\psi_m\
        e^{-S[\bar\psi,\psi]},\\
        S[\bar\psi,\psi]
        =\sum_{m=1}^M\bar\psi_m\left[
            (\psi_m-\psi_{m-1})
            +\frac{\beta}{M}(\epsilon-\mu)\bar\psi_m\psi_{m-1}
            \right].
    \end{gathered}
\end{equation}
And consequently, the Green's function is transformed as
\begin{equation}
    \begin{gathered}
        G_0(\mathrm i\omega_n)=\frac{1}{\mathrm i\omega_n+\mu-\epsilon}
        \to\frac{1}{\mathrm i\tilde\omega_n+\mu-\epsilon},\\
        \mathrm i\tilde\omega_n=-\frac{M}{\beta}\left(
        e^{-\mathrm i\frac{(2n+1)\pi}{M}}-1
        \right).
    \end{gathered}
\end{equation}
It is easy to see that when $M\to\infty$ or $|n|\ll M$,
the discretization frequencies $\tilde\omega_n\to\omega_n$.
For any given observable $O$,
turning out to be $O(M)$ when $M$ is finite,
there is a systematic error of order $o(1/M)$.
To minimize this error as much as possible,
we believe that the following expansion would be obeyed
when $M$ is sufficiently large to satisfy condition
\begin{equation}
    O(M)=O(\infty)
    +\frac{a_1}{M}+\frac{a_2}{M^2}+o(1/M^3).
\end{equation}
We take $M$ to be $1000, 1200, 1400, 1600, 1800$, and $2000$,
and by fitting the expansion,
we obtain $O(\infty)$ as the estimate for $O$.

The \textit{PCGA Green's function $G$}
is calculated by Eqs.~(\ref{eq:self-energy},\ref{eq:dyson pcga}).
To be more specific,
suppose $n^A_\uparrow=n^B_\downarrow=\frac{1}{2}(n+m)$,
$n^B_\uparrow=n^A_\downarrow=\frac{1}{2}(n-m)$,
the Hartree-Fock Green's function in matrix form is
\begin{gather}
    \left[g^{-1}_{\uparrow\uparrow}(\mathrm i\omega_n,\vec k)\right]^{AB}
    =\left[g^{-1}_{\downarrow\downarrow}(\mathrm i\omega_n,\vec k)\right]^{AB}
    =-\epsilon_2(\vec k)e^{+\mathrm i\varphi(\vec k)}, \\
    \left[g^{-1}_{\uparrow\uparrow}(\mathrm i\omega_n,\vec k)\right]^{BA}
    =\left[g^{-1}_{\downarrow\downarrow}(\mathrm i\omega_n,\vec k)\right]^{BA}
    =-\epsilon_2(\vec k)e^{-\mathrm i\varphi(\vec k)}, \\
    \left[g^{-1}_{\uparrow\uparrow}(\mathrm i\omega_n,\vec k)\right]^{AA}
    =\left[g^{-1}_{\downarrow\downarrow}(\mathrm i\omega_n,\vec k)\right]^{BB}
    =\mathrm i\omega_n+\mu-\frac{U}{2}(n-m)
    -\epsilon_1(\vec k),\\
    \left[g^{-1}_{\uparrow\uparrow}(\mathrm i\omega_n,\vec k)\right]^{BB}
    =\left[g^{-1}_{\downarrow\downarrow}(\mathrm i\omega_n,\vec k)\right]^{AA}
    =\mathrm i\omega_n+\mu-\frac{U}{2}(n+m)
    -\epsilon_1(\vec k).
\end{gather}
Adopting the shorthand notation $k=(\mathrm i\omega_n,\vec k)$,
the PCGA self-energy in matrix form is
\begin{gather}
    \Sigma^{AB}_{\uparrow\uparrow}(k)
    =-\frac{U^2}{\beta N^2}\sum_{q_1,q_2}
    g^{BA}_{\downarrow\downarrow}(q_1+q_2-k)
    g^{AB}_{\uparrow\uparrow}(q_1)
    g^{AB}_{\downarrow\downarrow}(q_2),\\
    \Sigma^{BA}_{\uparrow\uparrow}(k)
    =-\frac{U^2}{\beta N^2}\sum_{q_1,q_2}
    g^{AB}_{\downarrow\downarrow}(q_1+q_2-k)
    g^{BA}_{\uparrow\uparrow}(q_1)
    g^{BA}_{\downarrow\downarrow}(q_2),\\
    \Sigma^{AA}_{\uparrow\uparrow}(k)
    =-\frac{U^2}{\beta N^2}\sum_{q_1,q_2}
    g^{AA}_{\downarrow\downarrow}(q_1+q_2-k)
    g^{AA}_{\uparrow\uparrow}(q_1)
    g^{AA}_{\downarrow\downarrow}(q_2),\\
    \Sigma^{BB}_{\uparrow\uparrow}(k)
    =-\frac{U^2}{\beta N^2}\sum_{q_1,q_2}
    g^{BB}_{\downarrow\downarrow}(q_1+q_2-k)
    g^{BB}_{\uparrow\uparrow}(q_1)
    g^{BB}_{\downarrow\downarrow}(q_2),
\end{gather}
Here we only present the $\uparrow\uparrow$ component.
The reason is two-fold:
first, the other spin components have simple numerical relationships with the $\uparrow\uparrow$ component;
second, the subsequent calculations do not require the use of other components.
The PCGA Green's function $G$ obeys the Dyson equation
in matrix form
\begin{equation}
    G^{-1}_{\uparrow\uparrow}(\mathrm i\omega_n,\vec k)
    =g^{-1}_{\uparrow\uparrow}(\mathrm i\omega_n,\vec k)
    -\Sigma_{\uparrow\uparrow}(\mathrm i\omega_n,\vec k).
\end{equation}
The other spin components are as follows: $G_{\uparrow\downarrow}=G_{\downarrow\uparrow}=0$.
Also,
$G_{\downarrow\downarrow}^{AB}=G_{\uparrow\uparrow}^{AB}$,
$G_{\downarrow\downarrow}^{BA}=G_{\uparrow\uparrow}^{BA}$,
$G_{\downarrow\downarrow}^{AA}=G_{\uparrow\uparrow}^{BB}$,
and $G_{\downarrow\downarrow}^{BB}=G_{\uparrow\uparrow}^{AA}$,
These relationships are the same as those between the Hartree-Fock Green's function $g$.

\textit{The correlation functions $\Pi$} are calculated by
    {\small
        \begin{equation}
            \begin{aligned}
                \Pi^{xx}(\mathrm i\omega_n, \vec k)
                = & -\frac{2}{\beta N}\sum_{\vec q}\sum_{\mathrm i\nu_n}
                \sum_{\sigma_1\sigma_2}
                \mathrm{tr}\left[
                    \underline\gamma^{x}(\vec q,\vec k)
                    G_{\sigma_1\sigma_2}(\mathrm i\nu_n,\vec q)
                    \underline\gamma^{x}(\vec k+\vec q,-\vec k)
                    G_{\sigma_2\sigma_1}(\mathrm i\omega_n+\mathrm i\nu_n,\vec k+\vec q)
                \right]                                                              \\
                \Pi^{yxy}(\mathrm i\omega_n,\vec k)
                = & \Pi_1(\mathrm i\omega_n,\vec k)
                +\Pi_2(\mathrm i\omega_n,\vec k)
                +\Pi_3(\mathrm i\omega_n,\vec k)                                     \\
                \Pi_1(\mathrm i\omega_n, \vec k)
                = & -\frac{2}{\beta N}\sum_{\vec q}\sum_{\mathrm i\nu_n}
                \sum_{\sigma_1\sigma_2\sigma_3}
                \mathrm{tr}\left[
                    \underline\gamma^{y}(\vec q,\vec k)
                    G_{\sigma_1\sigma_2}(\mathrm i\nu_n,\vec q)
                    \underline\gamma^{x}(\vec q+\vec k,-\vec k)
                G_{\sigma_2\sigma_3}(\mathrm i\nu_n,\vec q+\vec k)\right.            \\
                  & \times\left.\underline\gamma^{y}(\vec q+\vec k,\vec 0)
                    G_{\sigma_3\sigma_1}(\mathrm i\nu_n+\mathrm i\omega_n,\vec q+\vec k)
                \right]                                                              \\
                \Pi_2(\mathrm i\omega_n, \vec k)
                = & -\frac{2}{\beta N}\sum_{\vec q}\sum_{\mathrm i\nu_n}
                \sum_{\sigma_1\sigma_2\sigma_3}
                \mathrm{tr}\left[
                    \underline\gamma^{y}(\vec q,\vec k)
                    G_{\sigma_1\sigma_2}(\mathrm i\nu_n,\vec q)
                    \underline\gamma^{y}(\vec q+\vec k,\vec 0)
                G_{\sigma_2\sigma_3}(\mathrm i\nu_n+\mathrm i\omega_n,\vec q)\right. \\
                  & \times\left.\underline\gamma^{x}(\vec q+\vec k,-\vec k)
                    G_{\sigma_3\sigma_1}(\mathrm i\nu_n+\mathrm i\omega_n,\vec q+\vec k)
                \right]                                                              \\
                \Pi_3(\mathrm i\omega_n, \vec k)
                = & -\frac{2}{\beta N}\sum_{\vec q}\sum_{\mathrm i\nu_n}
                \sum_{\sigma_1\sigma_2}
                \mathrm{tr}\left[
                    \underline\gamma^{y}(\vec q,\vec k)
                    G_{\sigma_1\sigma_2}(\mathrm i\nu_n,\vec q)
                    \underline\gamma^{xy}_B(\vec k+\vec q,-\vec k)
                    G_{\sigma_2\sigma_1}(\mathrm i\omega_n+\mathrm i\nu_n,\vec k+\vec q)
                \right]                                                              \\
            \end{aligned}
            \label{eq:pi green}
        \end{equation}
    }
where $\mathrm{tr}$ take the trace refers specifically to the operations performed on the $AB$-sublattice.
The $\underline\gamma$ matrices are defined as
\begin{equation}
    \underline\gamma(\vec k,\vec q)
    =\frac{1}{2}\left(\underline\gamma(\vec k)+\underline\gamma(\vec k+\vec q)\right)
    \simeq\underline{\gamma}(\vec k+\vec q/2),
\end{equation}
\begin{equation}
    \underline\gamma^x(\vec k)=\begin{bmatrix}
        \partial_{k_x}\epsilon_1(\vec k)                              &
        e^{+\mathrm i\varphi(\vec k)}\partial_{k_x}\epsilon_2(\vec k)   \\
        e^{-\mathrm i\varphi(\vec k)}\partial_{k_x}\epsilon_2(\vec k) &
        \partial_{k_x}\epsilon_1(\vec k)
    \end{bmatrix},
\end{equation}
\begin{equation}
    \underline\gamma^{xy}_B(\vec k)=\begin{bmatrix}
        \partial_{k_x}\partial_{k_y}\epsilon_1(\vec k)                              &
        e^{+\mathrm i\varphi(\vec k)}\partial_{k_x}\partial_{k_y}\epsilon_2(\vec k)   \\
        e^{-\mathrm i\varphi(\vec k)}\partial_{k_x}\partial_{k_y}\epsilon_2(\vec k) &
        \partial_{k_x}\partial_{k_y}\epsilon_1(\vec k)
    \end{bmatrix},
\end{equation}

Finally, \textit{Hall number $n_H$} is computed by Eq.~(\ref{eq:Rh})
\begin{equation}
    n_H(\mathrm i\omega_n)=\left[
        \frac{\mathrm i\omega_n}{[\Pi^{xx}(\mathrm i\omega_n,\vec k)]^2}
        \frac{\partial\Pi^{yxy}(\mathrm i\omega_n,\vec k)}{\partial k_x}.
        \right]^{-1}_{\vec k=0}
\end{equation}
The partial derivative with respect to $k_x$
is approximated by performing a polynomial fit that includes only odd powers of
$k_x$ using the points $\vec k=(k_x,0)$ with the smallest values of $k_x$.
The resulting expression for $n_H$
is then taken as the smallest positive frequency
\begin{equation}
    n_H\simeq n_H(\mathrm i\frac{\pi}{\beta}).
\end{equation}
\section{\label{ap:next_order_self_energy}
  Perturbation theory upon Hartree-Fock approximation
 }
Starting from the action of the equilibrium field theory path integral $S[\bar\psi,\psi]$
\begin{equation}
    \begin{aligned}
        S[\bar\psi,\psi]
        = & \int\mathrm dr_1\mathrm dr_2\ \sum_{\sigma=\uparrow,\downarrow}
        \bar\psi_\sigma(r_1)\left[-G_0^{-1}(r_1,r_2)\right]\psi_\sigma(r_2) \\
          & +U\int\mathrm dr\ \bar\psi_\uparrow(r)\bar\psi_\downarrow(r)
        \psi_\downarrow(r)\psi_\uparrow(r),
    \end{aligned}
\end{equation}
and the Hartree-Fock action $S_\mathrm{MF}[\bar\psi,\psi]$
\begin{equation}
    \begin{aligned}
        S_\mathrm{MF}[\bar\psi,\psi]
        = & \int\mathrm dr_1\mathrm dr_2\ \sum_{\sigma=\uparrow,\downarrow}
        \bar\psi_\sigma(r_1)\left[-G_0^{-1}(r_1,r_2)\right]\psi_\sigma(r_2) \\
          & +U\int\mathrm dr\ n_{\mathrm{MF}\downarrow}(r)
        \bar\psi_\uparrow(r)\psi_\uparrow(r)                                \\
          & +U\int\mathrm dr\ n_{\mathrm{MF}\uparrow}(r)
        \bar\psi_\downarrow(r)\psi_\downarrow(r),
    \end{aligned}
\end{equation}
The difference between these two actions is identified as the perturbative part,
and we provide the perturbative expression for the Green's function.
\begin{equation}
    \begin{aligned}
        G_{\sigma_1\sigma_2}(r_1,r_2)
        = & \frac{1}{Z}\int D[\bar\psi,\psi]\
        \bar\psi_{\sigma_2}(r_2)\psi_{\sigma_1}(r_1)
        e^{-S[\bar\psi,\psi]}
        = \left\langle
        \bar\psi_{\sigma_2}(r_2)\psi_{\sigma_1}(r_1)
        e^{-S_\mathrm{int}[\bar\psi,\psi]}
        \right\rangle_c,                                                      \\
        S_\mathrm{int}[\bar\psi,\psi]
        = & U\int\mathrm dr\ \left[\bar\psi_\uparrow(r)\bar\psi_\downarrow(r)
        \psi_\downarrow(r)\psi_\uparrow(r)
        -n_{\mathrm{MF}\downarrow}
        \bar\psi_\uparrow(r)\psi_\uparrow(r)
        -n_{\mathrm{MF}\uparrow}
        \bar\psi_\downarrow(r)\psi_\downarrow(r)
        \right],
    \end{aligned}
\end{equation}
Here subscript $c$ denotes to \textit{connected} diagrams.
The lowest-order correction to the Green's function is zero,
which indicates the role of cancellation terms as
the subtraction of tadpole diagrams from
the one-particle irreducible (1PI) self-energy diagrams.
\begin{equation}
    \begin{aligned}
        G^{(1)}_{\upuparrows}(r_1,r_2)
        = & -U\int\mathrm dr_3\ \left\langle
        \bar\psi_\uparrow(r_2)\psi_\uparrow(r_1)
        \bar\psi_\uparrow(r_3)\bar\psi_\downarrow(r_3)
        \psi_\downarrow(r_3)\psi_\uparrow(r_3)
        \right\rangle_c                                        \\
          & +U\int\mathrm dr_3\ n_{\mathrm{MF}\downarrow}(r_3)
        \left\langle
        \bar\psi_\uparrow(r_2)\psi_\uparrow(r_1)
        \bar\psi_\uparrow(r_3)\psi_\uparrow(r_3)
        \right\rangle_c                                        \\
          & +U\int\mathrm dr_3\ n_{\mathrm{MF}\uparrow}(r_3)
        \left\langle
        \bar\psi_\uparrow(r_2)\psi_\uparrow(r_1)
        \bar\psi_\downarrow(r_3)\psi_\downarrow(r_3)
        \right\rangle_c                                        \\
        = & +U\int\mathrm dr_3\
        g_{\upuparrows}(r_1,r_3)
        g_{\downdownarrows}(r_3,r_3)
        g_{\upuparrows}(r_3,r_2)                               \\
          & -U\int\mathrm dr_3\ n_{\mathrm{MF}\downarrow}(r_3)
        g_{\upuparrows}(r_1,r_3)g_{\upuparrows}(r_3,r_2)=0.
    \end{aligned}
\end{equation}
The lowest non-trivial diagram is shown in Fig.~\ref{fig:ap d},
and its momentum space representation is precisely given by Eq.~(\ref{eq:self-energy}).
The next-order correction is also presented, with its momentum space form expressed as
\begin{equation}
    \begin{aligned}
        \Sigma^{(3)}_{\upuparrows}(k)
        = & -\frac{U^3}{(\beta N)^3}
        \int\mathrm dq_1\ g_{\downdownarrows}(k+q_1)\left[
            \int\mathrm dq_2\ g_{\uparrow\uparrow}(q_1+q_2)g_{\downarrow\downarrow}(q_2)
        \right]^2                                          \\
          & -\frac{U^3}{(\beta N)^3}\int\mathrm dq_1\left[
            \int\mathrm dq_2\ g_{\uparrow\uparrow}(q_1-q_2)g_{\downarrow\downarrow}(q_2)
            \right]^2
        g_{\downarrow\downarrow}(q_1-k)                    \\
    \end{aligned}
\end{equation}
The effect of $\Sigma^{(3)}$ is negligible when $U$ is small enough.
\begin{figure}
    \centering
    \begin{tikzpicture}[baseline=(in.base)]
        \begin{feynhand}
            \vertex (in) at (-2, 0) {};
            \vertex (out) at (2, 0) {};
            \vertex [dot] (p1) at (-1.2, 0) {};
            \vertex [dot] (p2) at (+1.2, 0) {};
            \propag [fermion] (in) to[edge label=$\uparrow$] (p1);
            \propag [fermion] (p1) to[edge label=$\uparrow$] (p2);
            \propag [fermion] (p2) to[edge label=$\uparrow$] (out);
            \propag [fermion] (p1) to[half left, edge label=$\downarrow$] (p2);
            \propag [fermion] (p2) to[half left, edge label=$\downarrow$] (p1);
        \end{feynhand}
    \end{tikzpicture}
    \quad
    \begin{tikzpicture}[baseline=(in.base)]
        \begin{feynhand}
            \vertex (in) at (-1.8, 0) {};
            \vertex (out) at (1.8, 0) {};
            \vertex [dot] (p1) at (-1.2, 0) {};
            \vertex [dot] (p2) at (+1.2, 0) {};
            \vertex [dot] (mid) at (0, 0) {};

            \propag [fermion] (in) to[edge label=$\uparrow$] (p1);
            \propag [fermion] (p1) to[edge label=$\uparrow$] (mid);
            \propag [fermion] (mid) to[edge label=$\uparrow$] (p2);
            \propag [fermion] (p2) to[edge label=$\uparrow$] (out);

            \propag [fermion] (p2) to[half left, edge label=$\downarrow$] (mid);
            \propag [fermion] (mid) to[half left, edge label=$\downarrow$] (p1);
            \propag [fermion] (p1) to[half left, edge label=$\downarrow$] (p2);
        \end{feynhand}
    \end{tikzpicture}
    \quad
    \begin{tikzpicture}[baseline=(in.base)]
        \begin{feynhand}
            \vertex (in) at (-1.8, 0) {};
            \vertex (out) at (1.8, 0) {};
            \vertex [dot] (p1) at (-1.2, 0) {};
            \vertex [dot] (p2) at (+1.2, 0) {};
            \vertex [dot] (mid) at (0, 0) {};

            \propag [fermion] (in) to[edge label=$\uparrow$] (p1);
            \propag [fermion] (p1) to[edge label=$\uparrow$] (mid);
            \propag [fermion] (mid) to[edge label=$\uparrow$] (p2);
            \propag [fermion] (p2) to[edge label=$\uparrow$] (out);

            \propag [fermion] (p1) to[half right, edge label=$\downarrow$] (mid);
            \propag [fermion] (mid) to[half right, edge label=$\downarrow$] (p2);
            \propag [fermion] (p2) to[half right, edge label=$\downarrow$] (p1);
        \end{feynhand}
    \end{tikzpicture}
    \caption{\label{fig:ap d}
        The Feynman diagrams for PCGA self-energy,
        labeled for $\Sigma_{\upuparrows}$.
        The lowest non-trival order is $U^2$ as the first diagram,
        and the next order includes contributions from other two diagrams.
    }
\end{figure}

\bibliography{apssamp}

\begin{thebibliography}{41}%
\makeatletter
\providecommand \@ifxundefined [1]{%
 \@ifx{#1\undefined}
}%
\providecommand \@ifnum [1]{%
 \ifnum #1\expandafter \@firstoftwo
 \else \expandafter \@secondoftwo
 \fi
}%
\providecommand \@ifx [1]{%
 \ifx #1\expandafter \@firstoftwo
 \else \expandafter \@secondoftwo
 \fi
}%
\providecommand \natexlab [1]{#1}%
\providecommand \enquote  [1]{``#1''}%
\providecommand \bibnamefont  [1]{#1}%
\providecommand \bibfnamefont [1]{#1}%
\providecommand \citenamefont [1]{#1}%
\providecommand \href@noop [0]{\@secondoftwo}%
\providecommand \href [0]{\begingroup \@sanitize@url \@href}%
\providecommand \@href[1]{\@@startlink{#1}\@@href}%
\providecommand \@@href[1]{\endgroup#1\@@endlink}%
\providecommand \@sanitize@url [0]{\catcode `\\12\catcode `\$12\catcode
  `\&12\catcode `\#12\catcode `\^12\catcode `\_12\catcode `\%12\relax}%
\providecommand \@@startlink[1]{}%
\providecommand \@@endlink[0]{}%
\providecommand \url  [0]{\begingroup\@sanitize@url \@url }%
\providecommand \@url [1]{\endgroup\@href {#1}{\urlprefix }}%
\providecommand \urlprefix  [0]{URL }%
\providecommand \Eprint [0]{\href }%
\providecommand \doibase [0]{https://doi.org/}%
\providecommand \selectlanguage [0]{\@gobble}%
\providecommand \bibinfo  [0]{\@secondoftwo}%
\providecommand \bibfield  [0]{\@secondoftwo}%
\providecommand \translation [1]{[#1]}%
\providecommand \BibitemOpen [0]{}%
\providecommand \bibitemStop [0]{}%
\providecommand \bibitemNoStop [0]{.\EOS\space}%
\providecommand \EOS [0]{\spacefactor3000\relax}%
\providecommand \BibitemShut  [1]{\csname bibitem#1\endcsname}%
\let\auto@bib@innerbib\@empty
\bibitem [{\citenamefont {Vedeneev}(2021)}]{Vedeneev_2021}%
  \BibitemOpen
  \bibfield  {author} {\bibinfo {author} {\bibfnamefont {S.~I.}\ \bibnamefont
  {Vedeneev}},\ }\bibfield  {title} {\bibinfo {title} {Pseudogap problem in
  high-temperature superconductors},\ }\href
  {https://doi.org/10.3367/UFNe.2020.12.038896} {\bibfield  {journal} {\bibinfo
   {journal} {Physics-Uspekhi}\ }\textbf {\bibinfo {volume} {64}},\ \bibinfo
  {pages} {890} (\bibinfo {year} {2021})}\BibitemShut {NoStop}%
\bibitem [{\citenamefont {Varma}(1997)}]{varma_1997}%
  \BibitemOpen
  \bibfield  {author} {\bibinfo {author} {\bibfnamefont {C.~M.}\ \bibnamefont
  {Varma}},\ }\bibfield  {title} {\bibinfo {title} {Non-fermi-liquid states and
  pairing instability of a general model of copper oxide metals},\ }\href
  {https://doi.org/10.1103/PhysRevB.55.14554} {\bibfield  {journal} {\bibinfo
  {journal} {Phys. Rev. B}\ }\textbf {\bibinfo {volume} {55}},\ \bibinfo
  {pages} {14554} (\bibinfo {year} {1997})}\BibitemShut {NoStop}%
\bibitem [{\citenamefont {Kivelson}\ \emph {et~al.}(1998)\citenamefont
  {Kivelson}, \citenamefont {Fradkin},\ and\ \citenamefont
  {Emery}}]{kivelson_1998}%
  \BibitemOpen
  \bibfield  {author} {\bibinfo {author} {\bibfnamefont {S.~A.}\ \bibnamefont
  {Kivelson}}, \bibinfo {author} {\bibfnamefont {E.}~\bibnamefont {Fradkin}},\
  and\ \bibinfo {author} {\bibfnamefont {V.~J.}\ \bibnamefont {Emery}},\
  }\bibfield  {title} {\bibinfo {title} {Electronic liquid-crystal phases of a
  doped mott insulator},\ }\href {https://doi.org/10.1038/31177} {\bibfield
  {journal} {\bibinfo  {journal} {Nature}\ ,\ \bibinfo {pages} {550}} (\bibinfo
  {year} {1998})}\BibitemShut {NoStop}%
\bibitem [{\citenamefont {Sedrakyan}\ and\ \citenamefont
  {Chubukov}(2010)}]{Sedrakyan_2010}%
  \BibitemOpen
  \bibfield  {author} {\bibinfo {author} {\bibfnamefont {T.~A.}\ \bibnamefont
  {Sedrakyan}}\ and\ \bibinfo {author} {\bibfnamefont {A.~V.}\ \bibnamefont
  {Chubukov}},\ }\bibfield  {title} {\bibinfo {title} {Pseudogap in underdoped
  cuprates and spin-density-wave fluctuations},\ }\href
  {https://doi.org/10.1103/PhysRevB.81.174536} {\bibfield  {journal} {\bibinfo
  {journal} {Phys. Rev. B}\ }\textbf {\bibinfo {volume} {81}},\ \bibinfo
  {pages} {174536} (\bibinfo {year} {2010})}\BibitemShut {NoStop}%
\bibitem [{\citenamefont {Wu}\ \emph {et~al.}(2019)\citenamefont {Wu},
  \citenamefont {Abanov}, \citenamefont {Wang},\ and\ \citenamefont
  {Chubukov}}]{Wu_2019}%
  \BibitemOpen
  \bibfield  {author} {\bibinfo {author} {\bibfnamefont {Y.-M.}\ \bibnamefont
  {Wu}}, \bibinfo {author} {\bibfnamefont {A.}~\bibnamefont {Abanov}}, \bibinfo
  {author} {\bibfnamefont {Y.}~\bibnamefont {Wang}},\ and\ \bibinfo {author}
  {\bibfnamefont {A.~V.}\ \bibnamefont {Chubukov}},\ }\bibfield  {title}
  {\bibinfo {title} {Special role of the first matsubara frequency for
  superconductivity near a quantum critical point: Nonlinear gap equation below
  ${T}_{c}$ and spectral properties in real frequencies},\ }\href
  {https://doi.org/10.1103/PhysRevB.99.144512} {\bibfield  {journal} {\bibinfo
  {journal} {Phys. Rev. B}\ }\textbf {\bibinfo {volume} {99}},\ \bibinfo
  {pages} {144512} (\bibinfo {year} {2019})}\BibitemShut {NoStop}%
\bibitem [{\citenamefont {H\"ucker}\ \emph {et~al.}(2014)\citenamefont
  {H\"ucker}, \citenamefont {Christensen}, \citenamefont {Holmes},
  \citenamefont {Blackburn}, \citenamefont {Forgan}, \citenamefont {Liang},
  \citenamefont {Bonn}, \citenamefont {Hardy}, \citenamefont {Gutowski},
  \citenamefont {Zimmermann}, \citenamefont {Hayden},\ and\ \citenamefont
  {Chang}}]{Hucker_2014}%
  \BibitemOpen
  \bibfield  {author} {\bibinfo {author} {\bibfnamefont {M.}~\bibnamefont
  {H\"ucker}}, \bibinfo {author} {\bibfnamefont {N.~B.}\ \bibnamefont
  {Christensen}}, \bibinfo {author} {\bibfnamefont {A.~T.}\ \bibnamefont
  {Holmes}}, \bibinfo {author} {\bibfnamefont {E.}~\bibnamefont {Blackburn}},
  \bibinfo {author} {\bibfnamefont {E.~M.}\ \bibnamefont {Forgan}}, \bibinfo
  {author} {\bibfnamefont {R.}~\bibnamefont {Liang}}, \bibinfo {author}
  {\bibfnamefont {D.~A.}\ \bibnamefont {Bonn}}, \bibinfo {author}
  {\bibfnamefont {W.~N.}\ \bibnamefont {Hardy}}, \bibinfo {author}
  {\bibfnamefont {O.}~\bibnamefont {Gutowski}}, \bibinfo {author}
  {\bibfnamefont {M.~v.}\ \bibnamefont {Zimmermann}}, \bibinfo {author}
  {\bibfnamefont {S.~M.}\ \bibnamefont {Hayden}},\ and\ \bibinfo {author}
  {\bibfnamefont {J.}~\bibnamefont {Chang}},\ }\bibfield  {title} {\bibinfo
  {title} {Competing charge, spin, and superconducting orders in underdoped
  {${\mathrm{YBa}}_{2}{\mathrm{Cu}}_{3}{\mathrm{O}}_{y}$}},\ }\href
  {https://doi.org/10.1103/PhysRevB.90.054514} {\bibfield  {journal} {\bibinfo
  {journal} {Phys. Rev. B}\ }\textbf {\bibinfo {volume} {90}},\ \bibinfo
  {pages} {054514} (\bibinfo {year} {2014})}\BibitemShut {NoStop}%
\bibitem [{\citenamefont {Wang}\ \emph {et~al.}(2015)\citenamefont {Wang},
  \citenamefont {Agterberg},\ and\ \citenamefont {Chubukov}}]{Wang_2015}%
  \BibitemOpen
  \bibfield  {author} {\bibinfo {author} {\bibfnamefont {Y.}~\bibnamefont
  {Wang}}, \bibinfo {author} {\bibfnamefont {D.~F.}\ \bibnamefont
  {Agterberg}},\ and\ \bibinfo {author} {\bibfnamefont {A.}~\bibnamefont
  {Chubukov}},\ }\bibfield  {title} {\bibinfo {title} {Coexistence of
  charge-density-wave and pair-density-wave orders in underdoped cuprates},\
  }\href {https://doi.org/10.1103/PhysRevLett.114.197001} {\bibfield  {journal}
  {\bibinfo  {journal} {Phys. Rev. Lett.}\ }\textbf {\bibinfo {volume} {114}},\
  \bibinfo {pages} {197001} (\bibinfo {year} {2015})}\BibitemShut {NoStop}%
\bibitem [{\citenamefont {Comin}\ \emph {et~al.}(2015)\citenamefont {Comin},
  \citenamefont {Sutarto}, \citenamefont {da~Silva~Neto}, \citenamefont
  {Chauviere}, \citenamefont {Liang}, \citenamefont {Hardy}, \citenamefont
  {Bonn}, \citenamefont {He}, \citenamefont {Sawatzky},\ and\ \citenamefont
  {Damascelli}}]{comin_2015}%
  \BibitemOpen
  \bibfield  {author} {\bibinfo {author} {\bibfnamefont {R.}~\bibnamefont
  {Comin}}, \bibinfo {author} {\bibfnamefont {R.}~\bibnamefont {Sutarto}},
  \bibinfo {author} {\bibfnamefont {E.~H.}\ \bibnamefont {da~Silva~Neto}},
  \bibinfo {author} {\bibfnamefont {L.}~\bibnamefont {Chauviere}}, \bibinfo
  {author} {\bibfnamefont {R.}~\bibnamefont {Liang}}, \bibinfo {author}
  {\bibfnamefont {W.~N.}\ \bibnamefont {Hardy}}, \bibinfo {author}
  {\bibfnamefont {D.~A.}\ \bibnamefont {Bonn}}, \bibinfo {author}
  {\bibfnamefont {F.}~\bibnamefont {He}}, \bibinfo {author} {\bibfnamefont
  {G.~A.}\ \bibnamefont {Sawatzky}},\ and\ \bibinfo {author} {\bibfnamefont
  {A.}~\bibnamefont {Damascelli}},\ }\bibfield  {title} {\bibinfo {title}
  {Broken translational and rotational symmetry via charge stripe order in
  underdoped {YBa$_{2}$Cu$_{3}$O$_{6+y}$}},\ }\href
  {https://doi.org/10.1126/science.1258399} {\bibfield  {journal} {\bibinfo
  {journal} {Science}\ }\textbf {\bibinfo {volume} {347}},\ \bibinfo {pages}
  {1335} (\bibinfo {year} {2015})}\BibitemShut {NoStop}%
\bibitem [{\citenamefont {Bal\'edent}\ \emph {et~al.}(2011)\citenamefont
  {Bal\'edent}, \citenamefont {Haug}, \citenamefont {Sidis}, \citenamefont
  {Hinkov}, \citenamefont {Lin},\ and\ \citenamefont
  {Bourges}}]{Baledent_2011}%
  \BibitemOpen
  \bibfield  {author} {\bibinfo {author} {\bibfnamefont {V.}~\bibnamefont
  {Bal\'edent}}, \bibinfo {author} {\bibfnamefont {D.}~\bibnamefont {Haug}},
  \bibinfo {author} {\bibfnamefont {Y.}~\bibnamefont {Sidis}}, \bibinfo
  {author} {\bibfnamefont {V.}~\bibnamefont {Hinkov}}, \bibinfo {author}
  {\bibfnamefont {C.~T.}\ \bibnamefont {Lin}},\ and\ \bibinfo {author}
  {\bibfnamefont {P.}~\bibnamefont {Bourges}},\ }\bibfield  {title} {\bibinfo
  {title} {Evidence for competing magnetic instabilities in underdoped
  {YBa${}_{2}$Cu${}_{3}$O${}_{6+x}$}},\ }\href
  {https://doi.org/10.1103/PhysRevB.83.104504} {\bibfield  {journal} {\bibinfo
  {journal} {Phys. Rev. B}\ }\textbf {\bibinfo {volume} {83}},\ \bibinfo
  {pages} {104504} (\bibinfo {year} {2011})}\BibitemShut {NoStop}%
\bibitem [{\citenamefont {Cyr-Choini\`ere}\ \emph {et~al.}(2015)\citenamefont
  {Cyr-Choini\`ere}, \citenamefont {Grissonnanche}, \citenamefont {Badoux},
  \citenamefont {Day}, \citenamefont {Bonn}, \citenamefont {Hardy},
  \citenamefont {Liang}, \citenamefont {Doiron-Leyraud},\ and\ \citenamefont
  {Taillefer}}]{Cyr_2015}%
  \BibitemOpen
  \bibfield  {author} {\bibinfo {author} {\bibfnamefont {O.}~\bibnamefont
  {Cyr-Choini\`ere}}, \bibinfo {author} {\bibfnamefont {G.}~\bibnamefont
  {Grissonnanche}}, \bibinfo {author} {\bibfnamefont {S.}~\bibnamefont
  {Badoux}}, \bibinfo {author} {\bibfnamefont {J.}~\bibnamefont {Day}},
  \bibinfo {author} {\bibfnamefont {D.~A.}\ \bibnamefont {Bonn}}, \bibinfo
  {author} {\bibfnamefont {W.~N.}\ \bibnamefont {Hardy}}, \bibinfo {author}
  {\bibfnamefont {R.}~\bibnamefont {Liang}}, \bibinfo {author} {\bibfnamefont
  {N.}~\bibnamefont {Doiron-Leyraud}},\ and\ \bibinfo {author} {\bibfnamefont
  {L.}~\bibnamefont {Taillefer}},\ }\bibfield  {title} {\bibinfo {title} {Two
  types of nematicity in the phase diagram of the cuprate superconductor
  {${\mathrm{YBa}}_{2}{\mathrm{Cu}}_{3}{\mathrm{O}}_{y}$}},\ }\href
  {https://doi.org/10.1103/PhysRevB.92.224502} {\bibfield  {journal} {\bibinfo
  {journal} {Phys. Rev. B}\ }\textbf {\bibinfo {volume} {92}},\ \bibinfo
  {pages} {224502} (\bibinfo {year} {2015})}\BibitemShut {NoStop}%
\bibitem [{\citenamefont {Badoux}\ \emph {et~al.}(2016)\citenamefont {Badoux},
  \citenamefont {Tabis}, \citenamefont {Laliberté}, \citenamefont
  {Grissonnanche}, \citenamefont {Vignolle}, \citenamefont {Vignolles},
  \citenamefont {Béard}, \citenamefont {Bonn}, \citenamefont {Hardy},
  \citenamefont {Liang}, \citenamefont {Doiron-Leyraud}, \citenamefont
  {Taillefer},\ and\ \citenamefont {Proust}}]{badoux_change_2016}%
  \BibitemOpen
  \bibfield  {author} {\bibinfo {author} {\bibfnamefont {S.}~\bibnamefont
  {Badoux}}, \bibinfo {author} {\bibfnamefont {W.}~\bibnamefont {Tabis}},
  \bibinfo {author} {\bibfnamefont {F.}~\bibnamefont {Laliberté}}, \bibinfo
  {author} {\bibfnamefont {G.}~\bibnamefont {Grissonnanche}}, \bibinfo {author}
  {\bibfnamefont {B.}~\bibnamefont {Vignolle}}, \bibinfo {author}
  {\bibfnamefont {D.}~\bibnamefont {Vignolles}}, \bibinfo {author}
  {\bibfnamefont {J.}~\bibnamefont {Béard}}, \bibinfo {author} {\bibfnamefont
  {D.~A.}\ \bibnamefont {Bonn}}, \bibinfo {author} {\bibfnamefont {W.~N.}\
  \bibnamefont {Hardy}}, \bibinfo {author} {\bibfnamefont {R.}~\bibnamefont
  {Liang}}, \bibinfo {author} {\bibfnamefont {N.}~\bibnamefont
  {Doiron-Leyraud}}, \bibinfo {author} {\bibfnamefont {L.}~\bibnamefont
  {Taillefer}},\ and\ \bibinfo {author} {\bibfnamefont {C.}~\bibnamefont
  {Proust}},\ }\bibfield  {title} {\bibinfo {title} {Change of carrier density
  at the pseudogap critical point of a cuprate superconductor},\ }\href
  {https://doi.org/10.1038/nature16983} {\bibfield  {journal} {\bibinfo
  {journal} {Nature}\ }\textbf {\bibinfo {volume} {531}},\ \bibinfo {pages}
  {210} (\bibinfo {year} {2016})}\BibitemShut {NoStop}%
\bibitem [{\citenamefont {Greene}\ \emph {et~al.}(2020)\citenamefont {Greene},
  \citenamefont {Mandal}, \citenamefont {Poniatowski},\ and\ \citenamefont
  {Sarkar}}]{greene_strange_2020}%
  \BibitemOpen
  \bibfield  {author} {\bibinfo {author} {\bibfnamefont {R.~L.}\ \bibnamefont
  {Greene}}, \bibinfo {author} {\bibfnamefont {P.~R.}\ \bibnamefont {Mandal}},
  \bibinfo {author} {\bibfnamefont {N.~R.}\ \bibnamefont {Poniatowski}},\ and\
  \bibinfo {author} {\bibfnamefont {T.}~\bibnamefont {Sarkar}},\ }\bibfield
  {title} {\bibinfo {title} {The {Strange} {Metal} {State} of the
  {Electron}-{Doped} {Cuprates}},\ }\href
  {https://doi.org/10.1146/annurev-conmatphys-031119-050558} {\bibfield
  {journal} {\bibinfo  {journal} {Annu. Rev. Condens. Matter Phys.}\ }\textbf
  {\bibinfo {volume} {11}},\ \bibinfo {pages} {213} (\bibinfo {year}
  {2020})}\BibitemShut {NoStop}%
\bibitem [{\citenamefont {Storey}(2016)}]{storey_hall_2016}%
  \BibitemOpen
  \bibfield  {author} {\bibinfo {author} {\bibfnamefont {J.~G.}\ \bibnamefont
  {Storey}},\ }\bibfield  {title} {\bibinfo {title} {Hall effect and {Fermi}
  surface reconstruction via electron pockets in the high-${T}_c$ cuprates},\
  }\href {https://doi.org/10.1209/0295-5075/113/27003} {\bibfield  {journal}
  {\bibinfo  {journal} {EPL (Europhysics Letters)}\ }\textbf {\bibinfo {volume}
  {113}},\ \bibinfo {pages} {27003} (\bibinfo {year} {2016})},\ \bibinfo {note}
  {publisher: IOP Publishing}\BibitemShut {NoStop}%
\bibitem [{\citenamefont {Eberlein}\ \emph {et~al.}(2016)\citenamefont
  {Eberlein}, \citenamefont {Metzner}, \citenamefont {Sachdev},\ and\
  \citenamefont {Yamase}}]{Eberlein_2016}%
  \BibitemOpen
  \bibfield  {author} {\bibinfo {author} {\bibfnamefont {A.}~\bibnamefont
  {Eberlein}}, \bibinfo {author} {\bibfnamefont {W.}~\bibnamefont {Metzner}},
  \bibinfo {author} {\bibfnamefont {S.}~\bibnamefont {Sachdev}},\ and\ \bibinfo
  {author} {\bibfnamefont {H.}~\bibnamefont {Yamase}},\ }\bibfield  {title}
  {\bibinfo {title} {Fermi surface reconstruction and drop in the hall number
  due to spiral antiferromagnetism in high-${T}_{c}$ cuprates},\ }\href
  {https://doi.org/10.1103/PhysRevLett.117.187001} {\bibfield  {journal}
  {\bibinfo  {journal} {Phys. Rev. Lett.}\ }\textbf {\bibinfo {volume} {117}},\
  \bibinfo {pages} {187001} (\bibinfo {year} {2016})}\BibitemShut {NoStop}%
\bibitem [{\citenamefont {Mitscherling}\ and\ \citenamefont
  {Metzner}(2018)}]{mitscherling_longitudinal_2018}%
  \BibitemOpen
  \bibfield  {author} {\bibinfo {author} {\bibfnamefont {J.}~\bibnamefont
  {Mitscherling}}\ and\ \bibinfo {author} {\bibfnamefont {W.}~\bibnamefont
  {Metzner}},\ }\bibfield  {title} {\bibinfo {title} {Longitudinal conductivity
  and {Hall} coefficient in two-dimensional metals with spiral magnetic
  order},\ }\href {https://doi.org/10.1103/PhysRevB.98.195126} {\bibfield
  {journal} {\bibinfo  {journal} {Phys. Rev. B}\ }\textbf {\bibinfo {volume}
  {98}},\ \bibinfo {pages} {195126} (\bibinfo {year} {2018})}\BibitemShut
  {NoStop}%
\bibitem [{\citenamefont {Voruganti}\ \emph {et~al.}(1992)\citenamefont
  {Voruganti}, \citenamefont {Golubentsev},\ and\ \citenamefont
  {John}}]{Voruganti_conductivity_1992}%
  \BibitemOpen
  \bibfield  {author} {\bibinfo {author} {\bibfnamefont {P.}~\bibnamefont
  {Voruganti}}, \bibinfo {author} {\bibfnamefont {A.}~\bibnamefont
  {Golubentsev}},\ and\ \bibinfo {author} {\bibfnamefont {S.}~\bibnamefont
  {John}},\ }\bibfield  {title} {\bibinfo {title} {Conductivity and hall effect
  in the two-dimensional hubbard model},\ }\href
  {https://doi.org/10.1103/PhysRevB.45.13945} {\bibfield  {journal} {\bibinfo
  {journal} {Phys. Rev. B}\ }\textbf {\bibinfo {volume} {45}},\ \bibinfo
  {pages} {13945} (\bibinfo {year} {1992})}\BibitemShut {NoStop}%
\bibitem [{\citenamefont {Andersen}\ \emph {et~al.}(1995)\citenamefont
  {Andersen}, \citenamefont {Liechtenstein}, \citenamefont {Jepsen},\ and\
  \citenamefont {Paulsen}}]{andersen_lda_nodate}%
  \BibitemOpen
  \bibfield  {author} {\bibinfo {author} {\bibfnamefont {O.~K.}\ \bibnamefont
  {Andersen}}, \bibinfo {author} {\bibfnamefont {A.~I.}\ \bibnamefont
  {Liechtenstein}}, \bibinfo {author} {\bibfnamefont {O.}~\bibnamefont
  {Jepsen}},\ and\ \bibinfo {author} {\bibfnamefont {E.}~\bibnamefont
  {Paulsen}},\ }\bibfield  {title} {\bibinfo {title} {{LDA} energy bands,
  low-energy {Hamiltonians}, $t'$, $t''$, $t_{\bot} (\mathbf{k})$, and
  ${J}_{\bot}$.},\ }\href
  {https://doi.org/https://doi.org/10.1016/0022-3697(95)00269-3} {\bibfield
  {journal} {\bibinfo  {journal} {J. Phys. Chem. Solids}\ }\textbf {\bibinfo
  {volume} {56}},\ \bibinfo {pages} {1573} (\bibinfo {year}
  {1995})}\BibitemShut {NoStop}%
\bibitem [{\citenamefont {Pavarini}\ \emph {et~al.}(2001)\citenamefont
  {Pavarini}, \citenamefont {Dasgupta}, \citenamefont {Saha-Dasgupta},
  \citenamefont {Jepsen},\ and\ \citenamefont
  {Andersen}}]{pavarini_band-structure_2001}%
  \BibitemOpen
  \bibfield  {author} {\bibinfo {author} {\bibfnamefont {E.}~\bibnamefont
  {Pavarini}}, \bibinfo {author} {\bibfnamefont {I.}~\bibnamefont {Dasgupta}},
  \bibinfo {author} {\bibfnamefont {T.}~\bibnamefont {Saha-Dasgupta}}, \bibinfo
  {author} {\bibfnamefont {O.}~\bibnamefont {Jepsen}},\ and\ \bibinfo {author}
  {\bibfnamefont {O.~K.}\ \bibnamefont {Andersen}},\ }\bibfield  {title}
  {\bibinfo {title} {Band-structure trend in hole-doped cuprates and
  correlation with ${\mathit{t}}_{\mathit{c}\mathrm{max}}$},\ }\href
  {https://doi.org/10.1103/PhysRevLett.87.047003} {\bibfield  {journal}
  {\bibinfo  {journal} {Phys. Rev. Lett.}\ }\textbf {\bibinfo {volume} {87}},\
  \bibinfo {pages} {047003} (\bibinfo {year} {2001})}\BibitemShut {NoStop}%
\bibitem [{\citenamefont {Mor\'ee}\ \emph {et~al.}(2022)\citenamefont
  {Mor\'ee}, \citenamefont {Hirayama}, \citenamefont {Schmid}, \citenamefont
  {Yamaji},\ and\ \citenamefont {Imada}}]{Imada_ab_initio_2022}%
  \BibitemOpen
  \bibfield  {author} {\bibinfo {author} {\bibfnamefont {J.-B.}\ \bibnamefont
  {Mor\'ee}}, \bibinfo {author} {\bibfnamefont {M.}~\bibnamefont {Hirayama}},
  \bibinfo {author} {\bibfnamefont {M.~T.}\ \bibnamefont {Schmid}}, \bibinfo
  {author} {\bibfnamefont {Y.}~\bibnamefont {Yamaji}},\ and\ \bibinfo {author}
  {\bibfnamefont {M.}~\bibnamefont {Imada}},\ }\bibfield  {title} {\bibinfo
  {title} {Ab initio low-energy effective hamiltonians for the high-temperature
  superconducting cuprates
  {${\mathrm{Bi}}_{2}{\mathrm{Sr}}_{2}{\mathrm{CuO}}_{6},$
  ${\mathrm{Bi}}_{2}{\mathrm{Sr}}_{2}{\mathrm{CaCu}}_{2}{\mathrm{O}}_{8},$
  ${\mathrm{HgBa}}_{2}{\mathrm{CuO}}_{4},$ and ${\mathrm{CaCuO}}_{2}$}},\
  }\href {https://doi.org/10.1103/PhysRevB.106.235150} {\bibfield  {journal}
  {\bibinfo  {journal} {Phys. Rev. B}\ }\textbf {\bibinfo {volume} {106}},\
  \bibinfo {pages} {235150} (\bibinfo {year} {2022})}\BibitemShut {NoStop}%
\bibitem [{\citenamefont {Shen}\ and\ \citenamefont
  {Dessau}(1995)}]{Shen_electronic_1995}%
  \BibitemOpen
  \bibfield  {author} {\bibinfo {author} {\bibfnamefont {Z.-X.}\ \bibnamefont
  {Shen}}\ and\ \bibinfo {author} {\bibfnamefont {D.}~\bibnamefont {Dessau}},\
  }\bibfield  {title} {\bibinfo {title} {Electronic structure and photoemission
  studies of late transition-metal oxides — mott insulators and
  high-temperature superconductors},\ }\href
  {https://doi.org/https://doi.org/10.1016/0370-1573(95)80001-A} {\bibfield
  {journal} {\bibinfo  {journal} {Physics Reports}\ }\textbf {\bibinfo {volume}
  {253}},\ \bibinfo {pages} {1} (\bibinfo {year} {1995})}\BibitemShut {NoStop}%
\bibitem [{\citenamefont {Borisenko}\ \emph {et~al.}(2000)\citenamefont
  {Borisenko}, \citenamefont {Golden}, \citenamefont {Legner}, \citenamefont
  {Pichler}, \citenamefont {D\"urr}, \citenamefont {Knupfer}, \citenamefont
  {Fink}, \citenamefont {Yang}, \citenamefont {Abell},\ and\ \citenamefont
  {Berger}}]{Borisenko_joys_2000}%
  \BibitemOpen
  \bibfield  {author} {\bibinfo {author} {\bibfnamefont {S.~V.}\ \bibnamefont
  {Borisenko}}, \bibinfo {author} {\bibfnamefont {M.~S.}\ \bibnamefont
  {Golden}}, \bibinfo {author} {\bibfnamefont {S.}~\bibnamefont {Legner}},
  \bibinfo {author} {\bibfnamefont {T.}~\bibnamefont {Pichler}}, \bibinfo
  {author} {\bibfnamefont {C.}~\bibnamefont {D\"urr}}, \bibinfo {author}
  {\bibfnamefont {M.}~\bibnamefont {Knupfer}}, \bibinfo {author} {\bibfnamefont
  {J.}~\bibnamefont {Fink}}, \bibinfo {author} {\bibfnamefont {G.}~\bibnamefont
  {Yang}}, \bibinfo {author} {\bibfnamefont {S.}~\bibnamefont {Abell}},\ and\
  \bibinfo {author} {\bibfnamefont {H.}~\bibnamefont {Berger}},\ }\bibfield
  {title} {\bibinfo {title} {Joys and pitfalls of fermi surface mapping in
  {${\mathrm{Bi}}_{2}{\mathrm{Sr}}_{2}{\mathrm{CaCu}}_{2}\mathrm{O}_{8+\ensuremath{\delta}}$}
  using angle resolved photoemission},\ }\href
  {https://doi.org/10.1103/PhysRevLett.84.4453} {\bibfield  {journal} {\bibinfo
   {journal} {Phys. Rev. Lett.}\ }\textbf {\bibinfo {volume} {84}},\ \bibinfo
  {pages} {4453} (\bibinfo {year} {2000})}\BibitemShut {NoStop}%
\bibitem [{\citenamefont {Huang}\ \emph {et~al.}(2019)\citenamefont {Huang},
  \citenamefont {Sheppard}, \citenamefont {Moritz},\ and\ \citenamefont
  {Devereaux}}]{Huang_strange_2019}%
  \BibitemOpen
  \bibfield  {author} {\bibinfo {author} {\bibfnamefont {E.~W.}\ \bibnamefont
  {Huang}}, \bibinfo {author} {\bibfnamefont {R.}~\bibnamefont {Sheppard}},
  \bibinfo {author} {\bibfnamefont {B.}~\bibnamefont {Moritz}},\ and\ \bibinfo
  {author} {\bibfnamefont {T.~P.}\ \bibnamefont {Devereaux}},\ }\bibfield
  {title} {\bibinfo {title} {Strange metallicity in the doped hubbard model},\
  }\href {https://doi.org/10.1126/science.aau7063} {\bibfield  {journal}
  {\bibinfo  {journal} {Science}\ }\textbf {\bibinfo {volume} {366}},\ \bibinfo
  {pages} {987} (\bibinfo {year} {2019})}\BibitemShut {NoStop}%
\bibitem [{\citenamefont {Igoshev}\ \emph {et~al.}(2010)\citenamefont
  {Igoshev}, \citenamefont {Timirgazin}, \citenamefont {Katanin}, \citenamefont
  {Arzhnikov},\ and\ \citenamefont {Irkhin}}]{Igoshev_Incommensurate_2010}%
  \BibitemOpen
  \bibfield  {author} {\bibinfo {author} {\bibfnamefont {P.~A.}\ \bibnamefont
  {Igoshev}}, \bibinfo {author} {\bibfnamefont {M.~A.}\ \bibnamefont
  {Timirgazin}}, \bibinfo {author} {\bibfnamefont {A.~A.}\ \bibnamefont
  {Katanin}}, \bibinfo {author} {\bibfnamefont {A.~K.}\ \bibnamefont
  {Arzhnikov}},\ and\ \bibinfo {author} {\bibfnamefont {V.~Y.}\ \bibnamefont
  {Irkhin}},\ }\bibfield  {title} {\bibinfo {title} {Incommensurate magnetic
  order and phase separation in the two-dimensional hubbard model with nearest-
  and next-nearest-neighbor hopping},\ }\href
  {https://doi.org/10.1103/PhysRevB.81.094407} {\bibfield  {journal} {\bibinfo
  {journal} {Phys. Rev. B}\ }\textbf {\bibinfo {volume} {81}},\ \bibinfo
  {pages} {094407} (\bibinfo {year} {2010})}\BibitemShut {NoStop}%
\bibitem [{\citenamefont {Laughlin}(2014)}]{Laughlin_Hartree-Fock_2014}%
  \BibitemOpen
  \bibfield  {author} {\bibinfo {author} {\bibfnamefont {R.~B.}\ \bibnamefont
  {Laughlin}},\ }\bibfield  {title} {\bibinfo {title} {Hartree-fock computation
  of the high-${T}_{c}$ cuprate phase diagram},\ }\href
  {https://doi.org/10.1103/PhysRevB.89.035134} {\bibfield  {journal} {\bibinfo
  {journal} {Phys. Rev. B}\ }\textbf {\bibinfo {volume} {89}},\ \bibinfo
  {pages} {035134} (\bibinfo {year} {2014})}\BibitemShut {NoStop}%
\bibitem [{\citenamefont {Scholle}\ \emph {et~al.}(2023)\citenamefont
  {Scholle}, \citenamefont {Bonetti}, \citenamefont {Vilardi},\ and\
  \citenamefont {Metzner}}]{Scholle_comprehensive_2023}%
  \BibitemOpen
  \bibfield  {author} {\bibinfo {author} {\bibfnamefont {R.}~\bibnamefont
  {Scholle}}, \bibinfo {author} {\bibfnamefont {P.~M.}\ \bibnamefont
  {Bonetti}}, \bibinfo {author} {\bibfnamefont {D.}~\bibnamefont {Vilardi}},\
  and\ \bibinfo {author} {\bibfnamefont {W.}~\bibnamefont {Metzner}},\
  }\bibfield  {title} {\bibinfo {title} {Comprehensive mean-field analysis of
  magnetic and charge orders in the two-dimensional hubbard model},\ }\href
  {https://doi.org/10.1103/PhysRevB.108.035139} {\bibfield  {journal} {\bibinfo
   {journal} {Phys. Rev. B}\ }\textbf {\bibinfo {volume} {108}},\ \bibinfo
  {pages} {035139} (\bibinfo {year} {2023})}\BibitemShut {NoStop}%
\bibitem [{\citenamefont {Kao}\ \emph {et~al.}(2023)\citenamefont {Kao},
  \citenamefont {Li},\ and\ \citenamefont {Rosenstein}}]{Kao_Unified_2023}%
  \BibitemOpen
  \bibfield  {author} {\bibinfo {author} {\bibfnamefont {H.~C.}\ \bibnamefont
  {Kao}}, \bibinfo {author} {\bibfnamefont {D.}~\bibnamefont {Li}},\ and\
  \bibinfo {author} {\bibfnamefont {B.}~\bibnamefont {Rosenstein}},\ }\bibfield
   {title} {\bibinfo {title} {Unified intermediate coupling description of
  pseudogap and strange metal phases of cuprates},\ }\href
  {https://doi.org/10.1103/PhysRevB.107.054508} {\bibfield  {journal} {\bibinfo
   {journal} {Phys. Rev. B}\ }\textbf {\bibinfo {volume} {107}},\ \bibinfo
  {pages} {054508} (\bibinfo {year} {2023})}\BibitemShut {NoStop}%
\bibitem [{\citenamefont {Armitage}\ \emph {et~al.}(2003)\citenamefont
  {Armitage}, \citenamefont {Lu}, \citenamefont {Kim}, \citenamefont
  {Damascelli}, \citenamefont {Shen}, \citenamefont {Ronning}, \citenamefont
  {Feng}, \citenamefont {Bogdanov}, \citenamefont {Zhou}, \citenamefont {Yang},
  \citenamefont {Hussain}, \citenamefont {Mang}, \citenamefont {Kaneko},
  \citenamefont {Greven}, \citenamefont {Onose}, \citenamefont {Taguchi},
  \citenamefont {Tokura},\ and\ \citenamefont
  {Shen}}]{armitage_angle-resolved_2003}%
  \BibitemOpen
  \bibfield  {author} {\bibinfo {author} {\bibfnamefont {N.~P.}\ \bibnamefont
  {Armitage}}, \bibinfo {author} {\bibfnamefont {D.~H.}\ \bibnamefont {Lu}},
  \bibinfo {author} {\bibfnamefont {C.}~\bibnamefont {Kim}}, \bibinfo {author}
  {\bibfnamefont {A.}~\bibnamefont {Damascelli}}, \bibinfo {author}
  {\bibfnamefont {K.~M.}\ \bibnamefont {Shen}}, \bibinfo {author}
  {\bibfnamefont {F.}~\bibnamefont {Ronning}}, \bibinfo {author} {\bibfnamefont
  {D.~L.}\ \bibnamefont {Feng}}, \bibinfo {author} {\bibfnamefont
  {P.}~\bibnamefont {Bogdanov}}, \bibinfo {author} {\bibfnamefont {X.~J.}\
  \bibnamefont {Zhou}}, \bibinfo {author} {\bibfnamefont {W.~L.}\ \bibnamefont
  {Yang}}, \bibinfo {author} {\bibfnamefont {Z.}~\bibnamefont {Hussain}},
  \bibinfo {author} {\bibfnamefont {P.~K.}\ \bibnamefont {Mang}}, \bibinfo
  {author} {\bibfnamefont {N.}~\bibnamefont {Kaneko}}, \bibinfo {author}
  {\bibfnamefont {M.}~\bibnamefont {Greven}}, \bibinfo {author} {\bibfnamefont
  {Y.}~\bibnamefont {Onose}}, \bibinfo {author} {\bibfnamefont
  {Y.}~\bibnamefont {Taguchi}}, \bibinfo {author} {\bibfnamefont
  {Y.}~\bibnamefont {Tokura}},\ and\ \bibinfo {author} {\bibfnamefont {Z.-X.}\
  \bibnamefont {Shen}},\ }\bibfield  {title} {\bibinfo {title} {Angle-resolved
  photoemission spectral function analysis of the electron-doped cuprate
  {${\mathrm{Nd}}_{1.85}{\mathrm{Ce}}_{0.15}{\mathrm{CuO}}_{4}$}},\ }\href
  {https://doi.org/10.1103/PhysRevB.68.064517} {\bibfield  {journal} {\bibinfo
  {journal} {Phys. Rev. B}\ }\textbf {\bibinfo {volume} {68}},\ \bibinfo
  {pages} {064517} (\bibinfo {year} {2003})}\BibitemShut {NoStop}%
\bibitem [{\citenamefont {Matsui}\ \emph {et~al.}(2007)\citenamefont {Matsui},
  \citenamefont {Takahashi}, \citenamefont {Sato}, \citenamefont {Terashima},
  \citenamefont {Ding}, \citenamefont {Uefuji},\ and\ \citenamefont
  {Yamada}}]{matsui_evolution_2007}%
  \BibitemOpen
  \bibfield  {author} {\bibinfo {author} {\bibfnamefont {H.}~\bibnamefont
  {Matsui}}, \bibinfo {author} {\bibfnamefont {T.}~\bibnamefont {Takahashi}},
  \bibinfo {author} {\bibfnamefont {T.}~\bibnamefont {Sato}}, \bibinfo {author}
  {\bibfnamefont {K.}~\bibnamefont {Terashima}}, \bibinfo {author}
  {\bibfnamefont {H.}~\bibnamefont {Ding}}, \bibinfo {author} {\bibfnamefont
  {T.}~\bibnamefont {Uefuji}},\ and\ \bibinfo {author} {\bibfnamefont
  {K.}~\bibnamefont {Yamada}},\ }\bibfield  {title} {\bibinfo {title}
  {Evolution of the pseudogap across the magnet-superconductor phase boundary
  of
  {${\mathrm{Nd}}_{2\ensuremath{-}x}{\mathrm{Ce}}_{x}\mathrm{Cu}{\mathrm{O}}_{4}$}},\
  }\href {https://doi.org/10.1103/PhysRevB.75.224514} {\bibfield  {journal}
  {\bibinfo  {journal} {Phys. Rev. B}\ }\textbf {\bibinfo {volume} {75}},\
  \bibinfo {pages} {224514} (\bibinfo {year} {2007})}\BibitemShut {NoStop}%
\bibitem [{\citenamefont {Proust}\ and\ \citenamefont
  {Taillefer}(2019)}]{louis_remarkable_2019}%
  \BibitemOpen
  \bibfield  {author} {\bibinfo {author} {\bibfnamefont {C.}~\bibnamefont
  {Proust}}\ and\ \bibinfo {author} {\bibfnamefont {L.}~\bibnamefont
  {Taillefer}},\ }\bibfield  {title} {\bibinfo {title} {The remarkable
  underlying ground states of cuprate superconductors},\ }\href
  {https://doi.org/https://doi.org/10.1146/annurev-conmatphys-031218-013210}
  {\bibfield  {journal} {\bibinfo  {journal} {Annual Review of Condensed Matter
  Physics}\ }\textbf {\bibinfo {volume} {10}},\ \bibinfo {pages} {409}
  (\bibinfo {year} {2019})}\BibitemShut {NoStop}%
\bibitem [{\citenamefont {Gindikin}\ and\ \citenamefont
  {Chubukov}(2024)}]{FS_Gindikin_2024}%
  \BibitemOpen
  \bibfield  {author} {\bibinfo {author} {\bibfnamefont {Y.}~\bibnamefont
  {Gindikin}}\ and\ \bibinfo {author} {\bibfnamefont {A.~V.}\ \bibnamefont
  {Chubukov}},\ }\bibfield  {title} {\bibinfo {title} {Fermi surface geometry
  and optical conductivity of a two-dimensional electron gas near an
  ising-nematic quantum critical point},\ }\href
  {https://doi.org/10.1103/PhysRevB.109.115156} {\bibfield  {journal} {\bibinfo
   {journal} {Phys. Rev. B}\ }\textbf {\bibinfo {volume} {109}},\ \bibinfo
  {pages} {115156} (\bibinfo {year} {2024})}\BibitemShut {NoStop}%
\bibitem [{\citenamefont {Peierls}(1933)}]{peierls_zur_1933}%
  \BibitemOpen
  \bibfield  {author} {\bibinfo {author} {\bibfnamefont {R.}~\bibnamefont
  {Peierls}},\ }\bibfield  {title} {\bibinfo {title} {Zur {Theorie} des
  {Diamagnetismus} von {Leitungselektronen}},\ }\href
  {https://doi.org/10.1007/BF01342591} {\bibfield  {journal} {\bibinfo
  {journal} {Zeitschrift für Physik}\ }\textbf {\bibinfo {volume} {80}},\
  \bibinfo {pages} {763} (\bibinfo {year} {1933})}\BibitemShut {NoStop}%
\bibitem [{\citenamefont {Wannier}(1962)}]{wannier_dynamics_1962}%
  \BibitemOpen
  \bibfield  {author} {\bibinfo {author} {\bibfnamefont {G.~H.}\ \bibnamefont
  {Wannier}},\ }\bibfield  {title} {\bibinfo {title} {Dynamics of band
  electrons in electric and magnetic fields},\ }\href
  {https://doi.org/10.1103/RevModPhys.34.645} {\bibfield  {journal} {\bibinfo
  {journal} {Rev. Mod. Phys.}\ }\textbf {\bibinfo {volume} {34}},\ \bibinfo
  {pages} {645} (\bibinfo {year} {1962})}\BibitemShut {NoStop}%
\bibitem [{\citenamefont {Vu\ifmmode \check{c}\else \v{c}\fi{}i\ifmmode
  \check{c}\else \v{c}\fi{}evi\ifmmode~\acute{c}\else \'{c}\fi{}}\ and\
  \citenamefont {\ifmmode~\check{Z}\else
  \v{Z}\fi{}itko}(2021)}]{Vucice_electrical_2021}%
  \BibitemOpen
  \bibfield  {author} {\bibinfo {author} {\bibfnamefont {J.}~\bibnamefont
  {Vu\ifmmode \check{c}\else \v{c}\fi{}i\ifmmode \check{c}\else
  \v{c}\fi{}evi\ifmmode~\acute{c}\else \'{c}\fi{}}}\ and\ \bibinfo {author}
  {\bibfnamefont {R.}~\bibnamefont {\ifmmode~\check{Z}\else \v{Z}\fi{}itko}},\
  }\bibfield  {title} {\bibinfo {title} {Electrical conductivity in the hubbard
  model: Orbital effects of magnetic field},\ }\href
  {https://doi.org/10.1103/PhysRevB.104.205101} {\bibfield  {journal} {\bibinfo
   {journal} {Phys. Rev. B}\ }\textbf {\bibinfo {volume} {104}},\ \bibinfo
  {pages} {205101} (\bibinfo {year} {2021})}\BibitemShut {NoStop}%
\bibitem [{\citenamefont {Lederer}\ \emph {et~al.}(2017)\citenamefont
  {Lederer}, \citenamefont {Schattner}, \citenamefont {Berg},\ and\
  \citenamefont {Kivelson}}]{Samuel_Superconductivity_2017}%
  \BibitemOpen
  \bibfield  {author} {\bibinfo {author} {\bibfnamefont {S.}~\bibnamefont
  {Lederer}}, \bibinfo {author} {\bibfnamefont {Y.}~\bibnamefont {Schattner}},
  \bibinfo {author} {\bibfnamefont {E.}~\bibnamefont {Berg}},\ and\ \bibinfo
  {author} {\bibfnamefont {S.~A.}\ \bibnamefont {Kivelson}},\ }\bibfield
  {title} {\bibinfo {title} {Superconductivity and non-fermi liquid behavior
  near a nematic quantum critical point},\ }\href
  {https://doi.org/10.1073/pnas.1620651114} {\bibfield  {journal} {\bibinfo
  {journal} {Proceedings of the National Academy of Sciences}\ }\textbf
  {\bibinfo {volume} {114}},\ \bibinfo {pages} {4905} (\bibinfo {year}
  {2017})},\ \Eprint
  {https://arxiv.org/abs/https://www.pnas.org/doi/pdf/10.1073/pnas.1620651114}
  {https://www.pnas.org/doi/pdf/10.1073/pnas.1620651114} \BibitemShut {NoStop}%
\bibitem [{\citenamefont {Wang}\ \emph {et~al.}(2021)\citenamefont {Wang},
  \citenamefont {Ding}, \citenamefont {Moritz}, \citenamefont {Schattner},
  \citenamefont {Huang},\ and\ \citenamefont
  {Devereaux}}]{Huang_Numerical_2021}%
  \BibitemOpen
  \bibfield  {author} {\bibinfo {author} {\bibfnamefont {W.~O.}\ \bibnamefont
  {Wang}}, \bibinfo {author} {\bibfnamefont {J.~K.}\ \bibnamefont {Ding}},
  \bibinfo {author} {\bibfnamefont {B.}~\bibnamefont {Moritz}}, \bibinfo
  {author} {\bibfnamefont {Y.}~\bibnamefont {Schattner}}, \bibinfo {author}
  {\bibfnamefont {E.~W.}\ \bibnamefont {Huang}},\ and\ \bibinfo {author}
  {\bibfnamefont {T.~P.}\ \bibnamefont {Devereaux}},\ }\bibfield  {title}
  {\bibinfo {title} {Numerical approaches for calculating the low-field dc hall
  coefficient of the doped hubbard model},\ }\href
  {https://doi.org/10.1103/PhysRevResearch.3.033033} {\bibfield  {journal}
  {\bibinfo  {journal} {Phys. Rev. Res.}\ }\textbf {\bibinfo {volume} {3}},\
  \bibinfo {pages} {033033} (\bibinfo {year} {2021})}\BibitemShut {NoStop}%
\bibitem [{\citenamefont {Klebel-Knobloch}\ \emph {et~al.}(2023)\citenamefont
  {Klebel-Knobloch}, \citenamefont {Tabi{\'s}}, \citenamefont {Gala},
  \citenamefont {Bari{\v s}i{\'c}}, \citenamefont {Sunko},\ and\ \citenamefont
  {Bari{\v s}i{\'c}}}]{klebel-knobloch_transport_2023}%
  \BibitemOpen
  \bibfield  {author} {\bibinfo {author} {\bibfnamefont {B.}~\bibnamefont
  {Klebel-Knobloch}}, \bibinfo {author} {\bibfnamefont {W.}~\bibnamefont
  {Tabi{\'s}}}, \bibinfo {author} {\bibfnamefont {M.~A.}\ \bibnamefont {Gala}},
  \bibinfo {author} {\bibfnamefont {O.~S.}\ \bibnamefont {Bari{\v s}i{\'c}}},
  \bibinfo {author} {\bibfnamefont {D.~K.}\ \bibnamefont {Sunko}},\ and\
  \bibinfo {author} {\bibfnamefont {N.}~\bibnamefont {Bari{\v s}i{\'c}}},\
  }\bibfield  {title} {\bibinfo {title} {Transport properties and doping
  evolution of the {Fermi} surface in cuprates},\ }\href
  {https://doi.org/10.1038/s41598-023-39813-z} {\bibfield  {journal} {\bibinfo
  {journal} {Sci Rep}\ }\textbf {\bibinfo {volume} {13}},\ \bibinfo {pages}
  {13562} (\bibinfo {year} {2023})}\BibitemShut {NoStop}%
\bibitem [{\citenamefont {Dagan}\ and\ \citenamefont
  {Greene}(2016)}]{dagan_fermi_2016}%
  \BibitemOpen
  \bibfield  {author} {\bibinfo {author} {\bibfnamefont {Y.}~\bibnamefont
  {Dagan}}\ and\ \bibinfo {author} {\bibfnamefont {R.~L.}\ \bibnamefont
  {Greene}},\ }\href@noop {} {\bibinfo {title} {Fermi {Surface}
  {Reconstruction} in the {Electron}-doped {Cuprate}
  {Pr}$_{(2-x)}${Ce$_x$CuO$_4$}}} (\bibinfo {year} {2016}),\ \bibinfo {note}
  {arXiv:1612.01703 [cond-mat]}\BibitemShut {NoStop}%
\bibitem [{\citenamefont {Vishik}(2018)}]{vishik_photoemission_2018}%
  \BibitemOpen
  \bibfield  {author} {\bibinfo {author} {\bibfnamefont {I.~M.}\ \bibnamefont
  {Vishik}},\ }\bibfield  {title} {\bibinfo {title} {Photoemission perspective
  on pseudogap, superconducting fluctuations, and charge order in cuprates: a
  review of recent progress},\ }\href
  {https://doi.org/10.1088/1361-6633/aaba96} {\bibfield  {journal} {\bibinfo
  {journal} {Reports on Progress in Physics}\ }\textbf {\bibinfo {volume}
  {81}},\ \bibinfo {pages} {062501} (\bibinfo {year} {2018})}\BibitemShut
  {NoStop}%
\bibitem [{\citenamefont {Dong}\ \emph {et~al.}(2019)\citenamefont {Dong},
  \citenamefont {Chen},\ and\ \citenamefont {Gull}}]{dong_dynamical_2019}%
  \BibitemOpen
  \bibfield  {author} {\bibinfo {author} {\bibfnamefont {X.}~\bibnamefont
  {Dong}}, \bibinfo {author} {\bibfnamefont {X.}~\bibnamefont {Chen}},\ and\
  \bibinfo {author} {\bibfnamefont {E.}~\bibnamefont {Gull}},\ }\bibfield
  {title} {\bibinfo {title} {Dynamical charge susceptibility in the {Hubbard}
  model},\ }\href {https://doi.org/10.1103/PhysRevB.100.235107} {\bibfield
  {journal} {\bibinfo  {journal} {Physical Review B}\ }\textbf {\bibinfo
  {volume} {100}},\ \bibinfo {pages} {235107} (\bibinfo {year}
  {2019})}\BibitemShut {NoStop}%
\bibitem [{\citenamefont {Jevicki}(1977)}]{JEVICKI_1977}%
  \BibitemOpen
  \bibfield  {author} {\bibinfo {author} {\bibfnamefont {A.}~\bibnamefont
  {Jevicki}},\ }\bibfield  {title} {\bibinfo {title} {On the ground state and
  infrared divergences of goldstone bosons in two dimensions},\ }\href
  {https://doi.org/https://doi.org/10.1016/0370-2693(77)90229-5} {\bibfield
  {journal} {\bibinfo  {journal} {Physics Letters B}\ }\textbf {\bibinfo
  {volume} {71}},\ \bibinfo {pages} {327} (\bibinfo {year} {1977})}\BibitemShut
  {NoStop}%
\bibitem [{\citenamefont {Negele}\ and\ \citenamefont
  {Orland}(2018)}]{negele_quantum_2018}%
  \BibitemOpen
  \bibfield  {author} {\bibinfo {author} {\bibfnamefont {J.~W.}\ \bibnamefont
  {Negele}}\ and\ \bibinfo {author} {\bibfnamefont {H.}~\bibnamefont
  {Orland}},\ }\href@noop {} {\emph {\bibinfo {title} {Quantum many-particle
  systems}}},\ Advanced book classics\ (\bibinfo  {publisher} {CRC Press,
  Taylor \& Francis Group},\ \bibinfo {address} {Boca Raton London New York},\
  \bibinfo {year} {2018})\BibitemShut {NoStop}%
\end{thebibliography}%

\end{document}